%% file: usenixsecurity2026.tex
\documentclass[letterpaper,twocolumn,10pt]{article}
\usepackage{usenix}

\usepackage{tikz}
\usepackage{amsmath}\usepackage{algorithm} 
\usepackage{algpseudocode}
\usepackage{tikz} 
\usepackage{xspace}

\usepackage[table,xcdraw]{xcolor}
\usepackage{hhline}
\usepackage{booktabs}
\usepackage{makecell}
\usepackage{graphicx} 
\usepackage{siunitx}
\usepackage{lstlinebgrd}
\usepackage{enumitem}
\usepackage{amssymb}
\usepackage{threeparttable}
\usepackage{float}
\usepackage{xurl}

\usepackage{xcolor}
\definecolor{clr-background}{RGB}{255,255,255}
\definecolor{clr-text}{RGB}{0,0,0}
\definecolor{clr-string}{RGB}{163,21,21}
\definecolor{clr-namespace}{RGB}{0,0,0}
\definecolor{clr-preprocessor}{RGB}{128,128,128}
\definecolor{clr-keyword}{RGB}{0,0,255}
\definecolor{clr-type}{RGB}{43,145,175}
\definecolor{clr-variable}{RGB}{0,0,0}
\definecolor{clr-constant}{RGB}{111,0,138} 
\definecolor{clr-comment}{RGB}{0,128,0}

\usepackage{listings}
\lstdefinestyle{VS2017}{
    frame=tb,
    numbers=left,
    numbersep=-5pt,
    captionpos=b,
    breaklines=true,
    basicstyle=\scriptsize,
	backgroundcolor=\color{clr-background},
	stringstyle=\color{clr-string},
	identifierstyle=\color{clr-variable}, 
	commentstyle=\color{clr-comment},
	directivestyle=\color{clr-preprocessor}, 
	keywordstyle=\color{clr-type},
	keywordstyle={[2]\color{clr-constant}}, 
	tabsize=4
}

\newcommand{\myles}[1]{\textcolor{orange}{}}
\newcommand{\jorge}[1]{\textcolor{purple}{}}
\newcommand{\ardalan}[1]{\textcolor{cyan}{}}

\newcommand{\sname}{Patchlings\xspace}

\begin{document}

\date{}

\title{\Large \bf Patchlings: Safety-Preserving Flash-Based Hotpatching\\for Automotive Microcontrollers}

\author{
{\rm Yuxin "Myles" Liu}\\
UC Irvine
\and
{\rm Sekar Kulandaivel}\\
Robert Bosch LLC - Research \& Technology Center
\and
{\rm Ardalan Amiri Sani}\\
UC Irvine
\and
{\rm Jorge Guajardo}\\
Robert Bosch LLC - Research \& Technology Center
} 

\maketitle

\input{contents/abstract}

\input{contents/introduction}

\input{contents/background}

\input{contents/motivation}

\input{contents/overview}

\input{contents/compliance}

\input{contents/patch_table}

\input{contents/timing_aware}

\input{contents/persistency}

\input{contents/prototype_and_evaluation}

\input{contents/discussions}

\input{contents/relatedwork}

\input{contents/conclusion}



\input{contents/appendices}

\bibliographystyle{plainurl}
\bibliography{references}

\end{document}

%% file: contents/abstract.tex
\begin{abstract}


The increasing presence of software in modern automobiles has created a growing need to deliver software updates throughout a vehicle's entire lifespan.
Traditional update methods are slow and require months of re-validation to comply with stringent safety standards like ISO 26262.
Although hotpatching offers a path to faster updates, existing solutions for real-time embedded systems are unsuitable for the automotive domain: they overlook regulatory compliance, demand extensive safety validation, and lack support for the flash-based Execute-in-Place (XIP) architecture commonly used in automotive electronic control units (ECUs).



We introduce \sname, the first hotpatching framework designed for compliance, safety, and persistence in automotive systems. 
It fills the gap in applying hotpatching to automotive systems and fundamentally reduces the mean-time-to-mitigate (MTTM) for vulnerabilities and bugs.
We implement and evaluate a complete prototype of \sname on an automotive-grade hardware platform, \textit{NXP S32K148EVB}, with both FreeRTOS and Zephyr. 
Our results demonstrate low and deterministic overhead (e.g., 3.3 $\mu$s when a patch is applied), small firmware size increase (e.g., as low as 6.34\%), and successful patching of different types of real CVEs, proving its real-world applicability and effectiveness. 

\end{abstract}

%% file: contents/introduction.tex
\section{Introduction}
Today, Software-Defined Vehicles (SDVs)\ardalan{I think this is the first time this acronym is used in the paper. Spell it out.}\myles{Fixed. I was restructuring the paper and forgot to update this.} have become the dominant trend across nearly all types of modern automobiles~\cite{trend_of_sdv_1} and are expected to account for at least 90\% of the new-vehicle market by 2029~\cite{trend_of_sdv_0}. 
Software running in automotive systems typically includes advanced driver-assistance systems (ADAS), autonomous driving systems, electrification systems for electric vehicles (EVs), and entertainment and connectivity systems.
On average, vehicles contained about 200 million lines of code in 2020, doubling from 100 million in 2015~\cite{software_loc_in_automotive}, with this number projecting to grow to 650 million in the near future.
By comparison, a Boeing 787 Dreamliner passenger jet contains only 14 million lines of code~\cite{software_loc_in_automotive_1}. 
As the amount of software in vehicles significantly increases, the number and frequency of bugs (especially safety and security related ones) have also risen~\cite{software_bugs_in_new_vehicles_0, software_bugs_in_new_vehicles_1, software_bugs_in_new_vehicles_2}. 
In the U.S. market, software-related problems accounted for 46\% of all vehicle recalls in 2024, affecting a total of 13.4 million units, a substantial increase from 14.2\% in 2023~\cite{software_bugs_in_new_vehicles_0}. 

Although more vehicles now support over-the-air (OTA) updates, they still rely on conventional firmware updates that erase and reprogram the flash memory of ECUs. 
This process can be slow (e.g., causing extended system downtime), may introduce unsafe states (e.g., "bricking" an ECU during an interrupted update), and often demands months of re-validation due to strict regulatory requirements such as the Automotive Safety Integrity Level (ASIL) in ISO 26262~\cite{iso26262Road}), resulting in long mean-time-to-mitigate (MTTM) vulnerabilities. 

Hotpatching, which updates code in a running system without requiring a reboot, offers a promising way to reduce MTTM and has already been studied and deployed in many OSes (e.g., Linux, Android)~\cite{arnold2009ksplice, su2024lpah, wang2023pet, xu2020automatic}. 
Recent work has extended hotpatching to real-time embedded systems, including hardware-trampoline approaches~\cite{niesler2021hera}, VM-based patching~\cite{he2022rapidpatch}, IR-level transformation with automatic patch generation~\cite{salehi2024autopatch}, and code-shadowing-based secure hotpatching~\cite{kintsugi25}.
However, these systems primarily target general-purpose embedded/IoT platforms and do not satisfy key needs of mission-critical automotive ECUs. 
Specifically, they (i) overlook stringent ASIL-D constraints such as explicit control and data flow requirements~\cite{parasoft_how_to_satisfy_iso26262}, (ii) still require lengthy validation comparable to traditional updates (often lasting months~\cite{regression_test_icst_2023_wuersching, automotive_length_software_testing}), and (iii) rely solely on volatile memory (i.e., RAM) for storing and executing patches, which conflicts with the flash-based XIP architecture of automotive ECUs and their long-term persistence requirements~\cite{vehicle_xip_architecture, nor_flash_automotive_industry_size}.

In this paper, we aim to provide a hotpatching framework that achieves fast MTTM for critical bugs and vulnerabilities while adhering to the most stringent automotive safety requirements (i.e., ASIL-D) and remaining compatible with existing automotive ECUs' flash-based XIP architecture.
Our design is guided by three objectives: align with ISO 26262 ASIL-D constraints, bound the potential impact of patches to reduce validation effort, and support corruption-resilient flash-based patch storage. 

We introduce \sname, an end-to-end hotpatching system for automotive ECUs that integrates at the C source level using static trampolines (placeholders). 
To be ASIL-D compliance-ready, \sname uses only static memory, enforces single-entry/single-exit (SESE) functions, restricts pointer usage, avoids dynamic control transfers such as unconditional jumps and recursion, has deterministic behavior and performance, and keeps all control and data flows explicit at the source level.
It employs task-aware patch deployment and timing-aware buffer-based patch execution to reduce post-patching validation effort while preserving task deadlines. 
Finally, \sname supports storing and executing patches in flash memory using an ``append-only'' storage strategy, ensuring corruption-resilience and maintaining hard real-time behavior.

We implement \sname on an automotive-grade evaluation board (i.e., \textit{NXP S32K148}) and integrate it with FreeRTOS and Zephyr. 
Our evaluation shows that \sname increases firmware size by 6-7\%, incurs microsecond-scale overhead ($\approx$ 2.1 $\mu$s / 1.5 $\mu$s without a patch, $\approx$ 3.9 $\mu$s / 3.3 $\mu$s with a patch on FreeRTOS/Zephyr), and preserves the hard real-time properties required in safety-critical automotive ECUs (e.g., ``brake-by-wire'' systems).
We also demonstrate that \sname can hotpatch real-world CVEs in both RTOSes. 

Overall, this paper makes the following contributions:
\begin{itemize}[noitemsep, nolistsep]
    \item A general hotpatching approach that explicitly addresses ASIL-D safety constraints.\ardalan{Why is compliance one separate item and safety and persistence packed into another? They should be either all together in one item or they should come in three separate items.}\myles{I can do that. I was thinking that for this contribution list, I'm staying at pure high-level (not thinking about the exact challenges), but stay on the concept level. Do you think being more specific on the contributions is better?}
    \item Design of \sname, the first safety-preserving hotpatching system for deeply embedded, flash-based devices.
    \item A complete system prototype of \sname that runs on two popular automotive RTOSes.
    \item A thorough evaluation on automotive-grade hardware, demonstrating performance and practical applicability.
\end{itemize}

Although we design and build \sname with automotive systems in mind, we believe that our approach generalizes to other safety-critical domains, including aerospace, medical devices, and industrial control systems, where safety and persistence are also essential. 
We do not claim this as a contribution of this paper as many of those platforms are closed-source and we could not evaluate our approach on them.

%% file: contents/background.tex
\section{Background}

\subsection{Automotive Systems}
Modern vehicles operate as complex real-time embedded systems. 
Their software is typically managed by an RTOS, which schedules multiple concurrent tasks. 
Most of these tasks are periodic, and they all have a key metric called the Worst Case Execution Time (WCET), which is the maximum allowed execution time for that task to complete its work within each period. 
This WCET is usually determined through extensive static and dynamic timing analysis during the system design and validation phase. 
For safety reasons, all of these periodic tasks must adhere to their strict timing deadlines.

\subsection{Regulatory Landscape}


Automotive software is governed by stringent safety and security standards.
ISO 26262 defines a risk classification scheme called ASIL, ranging from ASIL A to ASIL D, with ASIL D being the most stringent, demanding practices such as deterministic behavior design~\cite{iso26262Road, parasoft_how_to_satisfy_iso26262}. 
More recently, UNECE R155 and R156 have mandated that manufacturers must manage vulnerabilities and software updates throughout the lifecycle of the vehicle~\cite{unece_r155, unece_r156}.
There are also guidance documents from NHTSA~\cite{nhtsa_guidance} that provide recommendations for the development and deployment of safe and secure software in SDVs. 
It should be noted that regulations for automotive software are still evolving, where existing regulations are expected to be updated and new regulations may be introduced or come into effect~\cite{iso26262_third_edition, china_gb_44495, china_gb_44496, iso_21448_2022}.

\subsection{Software Updates and Hotpatching}


Traditionally, updating firmware in embedded systems (e.g., automotive ECUs) has been a lengthy process that requires taking the system offline, reflashing the firmware, checking system integrity, and finally rebooting the system. 
Hotpatching, also known as Dynamic Software Updating (DSU), is a technique that allows for code to be updated at runtime without a system reboot. 
It often works by writing the patch code to a separate memory region and dynamically redirecting the execution flow. 

\subsection{Flash Memory in Automotive ECUs}

Most automotive ECUs use NOR flash memory as their primary non-volatile storage, which stores each bit in a cell and organizes cells into sectors. 
Its key advantage is the ability to support Execute-in-Place (XIP), where the CPU can execute code directly from flash without first copying it to RAM. 
Although reading from it is straight-forward (byte-addressable), writing to it is more complex and involves erasing the sector and programming the cells. 
In the erasing step, all the cells in the sector are erased to a predefined state (usually 1s). 
In the programming step, the cells are written with the new data by programming selected cells to 0s, and this process is done at the granularity of a bundle of cells at a time (usually 4, 8, or 16 cells). 
Both of these two steps are done by the flash memory controller, and it requires a certain amount of time to complete each step (with the erasing step being most time-consuming), where the flash bank becomes unavailable (for reading) during this time. 
This process is also susceptible to corruption if interrupted by a power loss, though a corruption can be detected by the flash memory controller's built-in Error Correction Code (ECC) mechanism and a bus error will be issued to the CPU.

%% file: contents/motivation.tex
\section{Motivation}
\label{sec:motivation}

\subsection{Timely Software Updates for ECUs}
Our focus in this paper is on automotive ECUs. 
With the electrification of modern vehicles, the number of ECUs per vehicle has increased~\cite{car_getting_more_mcu_0, car_getting_more_mcu_1}. 
A substantial number of ECUs remain powered even when the vehicle is shut off~\cite{car_off_but_mcu_still_on_0, car_off_but_mcu_still_on_1, car_off_but_mcu_still_on_2}.
For example, in an EV, the battery management system (BMS) continuously operates to monitor battery health and safety.
Software updates for ECUs are becoming more accessible (e.g., via over-the-air (OTA) updates). However, they still cannot reach all ECUs (especially those deeply embedded ones) due to both the complexity (e.g., behavior changes in one ECU can affect the entire system) and risk (e.g., interrupted firmware update can "brick" the ECU) associated with updating ECU firmware.

There have been numerous cases where updates caused failures or even "bricked" the target ECU (potentially rendering the vehicle unsafe or undrivable)~\cite{mache_door_locked_out, cybertruck_bricking_after_ota, rivian_bricking_during_software_update, vw_id4_bricking_after_ota, rivian_bricking_after_ota, jeep_bricking_after_ota}. 
These risks stem from widespread use of flash memory to host ECU firmware.
Traditional update methods typically erase and rewrite existing flash contents to install the new firmware, which can leave the system in an unsafe state if the process is interrupted (e.g., due to power loss). 
Although there are solutions to enable safe updates such as A/B partitioning~\cite{safe_ota_with_ab_partition, android_ab_update, mcuboot_with_zephyr, freertos_ota_update_stm32u5}, they either do not fully eliminate these risks or significantly increase hardware cost.

In recent years, hotpatching for real-time embedded systems has become an active research direction to address vulnerabilities without system downtime or re-programming the firmware~\cite{niesler2021hera, he2022rapidpatch, salehi2024autopatch, kintsugi25}. 
However, these approaches primarily target general-purpose IoT devices and thus are not designed with regulatory compliance in mind. 
Furthermore, none of the existing work can preserve functional or timing safety in automotive ECUs, and still require extensive system-wide validation (thus they cannot reduce the MTTM), not to mention their reliance on RAM for hosting patches (which is not practical for automotive). 
\ardalan{Another example that those three high-level benefits are not always mentioned in parallel. Here, safety and persistence are mentioned, but not compliance. This makes the storyline of the paper inconsistent.}
\myles{I was trying to use the safety aspect to cover both compliance and safety-preserving. I made it more clear now by spliting the sentence into two.}
\ardalan{This is still confusing to me. It seems like we were talking about safety before but now it's talking about compliance.} 
\myles{I further revised it to make it more clear. cannot preserve safety -> need extensive system-wide validation -> cannot reduce MTTM. }
We offer a more in-depth analysis of all related approaches in Section~\ref{sec:related_work}.

\subsection{Our Goals}
To address the gaps of prior work and enable emergency fixes for automotive ECUs, we set the following goals for a practical hotpatching framework that aims to reduce the MTTM:

\textbf{G1 - Compliance.}
The patching framework itself must be structured to be compliance-ready to automotive regulatory standards (e.g., ISO 26262), including explicit control and data flow, deterministic behavior, bi-directional traceability, static memory usage, etc.

\textbf{G2 - Safety preservation.}
The safety impact (i.e., changes in functional and timing behavior) of a patch needs to be bounded to a well-defined scope (e.g., a subset of tasks) by the patching framework itself. 
In this way, we can circumvent the need of time-consuming system-wide verification and validation after each patch, thus effectively reducing the MTTM.


\textbf{G3 - Persistent, corruption-resilient, real-time storage.}
Patches must be stored and executed from flash, survive reboots, and tolerate flash corruption (e.g., due to power loss) while maintaining the hard real-time property.

%% file: contents/overview.tex
\section{Overview}

\begin{figure*} 
    \centering
    \includegraphics[width=0.85\textwidth]{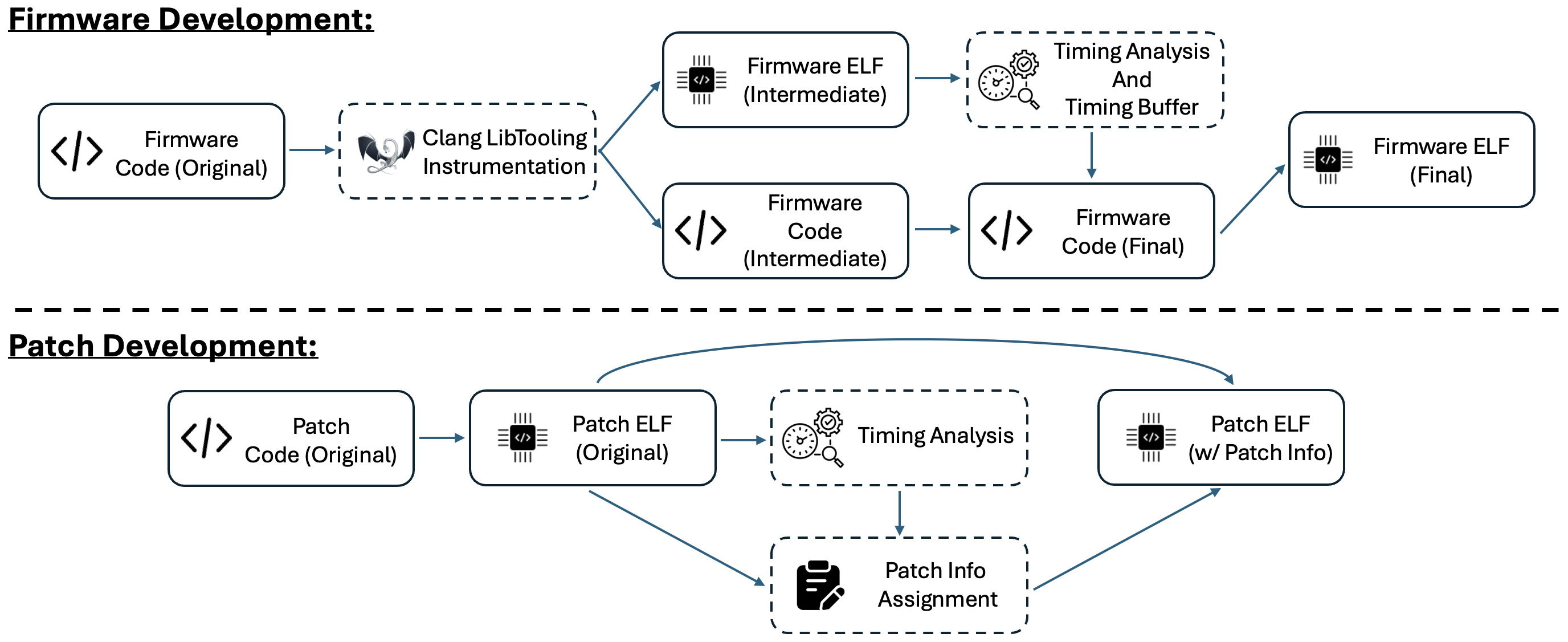}
    \caption{\em Overview of how \sname is integrated to both original firmware development and patch development stages.}
    \label{fig:design_overview}
\end{figure*}

\sname is carefully crafted to align with ASIL-D classification constraints, while significantly reducing the validation burden and enabling persistent hotpatching on flash-based automotive ECUs.
Concretely, it addresses the design goals in Section~\ref{sec:motivation} with the following solutions:

\textbf{S1 - ASIL-D-oriented design (addressing G1).}
\sname is designed to work at the C source code level, with deterministic behavior, explicit data flow or control flow, no unconditional jumps, no recursion, no dynamic memory allocation, restricted use of pointers, etc.
For example, all patching related functions are invoked via normal function calls, and patches can only keep or skip the entire original patched code section, avoiding hidden control flow. 
These choices are intended to make \sname compliance-ready with ASIL-D classification as defined in ISO 26262~\cite{iso26262Road, parasoft_how_to_satisfy_iso26262}.
We provide a comprehensive discussion on how \sname satisfies key ASIL-D requirements in Section~\ref{sec:compliance}.

\textbf{S2 - Task-aware patch deployment (addressing G2).}
\sname uses inlined placeholders and a patch dispatcher function that identify both the patched code location (via LR register) and the current task, and it maintains a per-patch task mask together with a global task bitmask.
This enables task-based patching: a patch is only activated when its target code section is invoked by designated task(s), bounding the patch's functional impact and thus the scope of functional verification and validation to the post-patching system.
We go into details about this task-based patch deployment process in Section~\ref{sec:patch_deployment}.

\textbf{S3 - Timing-aware patch execution (addressing G2).}
To preserve timing safety, \sname introduces per-task timing buffers, derived from the WCET of each task's periodic/sporadic code section.
Patches record their WCET in metadata (i.e., \texttt{patch\_info}), and our dispatcher deducts the patch WCET from the active task's timing buffer, while the remaining active task's timing buffer is emulated at the end of the current periodic/sporadic code section. 
Under the assumptions that WCETs and buffers are analyzed and configured correctly and buffers are not exhausted, \sname preserves task deadlines without requiring a full system-wide timing analysis after each patch deployment. 
This solution, combined with S2, upholds the system's functional and timing safety of the post-patching system, and significantly reduces the verification and validation effort, thus contributing to a fast MTTM. 
We discuss in details about how we preserve system timing behaviors in Section~\ref{sec:timing_awareness}. 

\textbf{S4 - Flash-based "append-only" patch storage (addressing G3).}
\sname stores both patch metadata and code in a reserved flash region using an "append-only" storage strategy along with a sector-sized recovery area.
New patches are written sequentially, and existing patches are never modified in place. 
Combined with a boot-time recovery procedure and a robust hardware software cooperative corruption detector, this design ensures that a valid \texttt{patch\_info} always points to valid patch code, limits corruption to a single sector, avoids flash erasures during normal operation, and supports persistent patches that survive reboots. 
Another benefit of this design is that it allows us to maintain the hard real-time property of the system (i.e., capability of running high-frequency critical tasks).
We elaborate on flash-based patching in Section~\ref{sec:patch_persistency}.


\textbf{\sname workflow.}
Figure~\ref{fig:design_overview} shows the workflow of \sname, where there are two phases: firmware development and patch development. 
In the firmware development phase, we expect both the \sname library and placeholders to be integrated into the original firmware. 
In the patch development phase, we expect the patch to be developed and configured properly for \sname to deploy it to the system. 
\myles{Is it better to leave the workflow here at the end of the overview section, or move it to the start of the overview section?}

%% file: contents/compliance.tex
\section{ASIL-D Compliance-Readiness}
\label{sec:compliance}

\sname is designed to be compliance-ready with the highest automotive safety classification, ASIL-D, meaning that its mechanisms and implementation choices are structured to align with key constraints and recommendations from the ASIL-D standard of ISO 26262~\cite{iso26262_third_edition, parasoft_how_to_satisfy_iso26262}. 
It should be noted that this does not automatically make \sname ASIL-D certified. 
ASIL-D certification is a rigorous process that is not just on the ECU itself, but is conducted on the entire top-to-bottom system, typically including sensors, ECUs, actuators, and even mechanical linkages. 
Therefore, we leave the actual certification process to Original Equipment Manufacturers (OEMs) and Tier-1 suppliers. 

To align with ASIL-D classification constraints, \sname adopts a deliberately conservative design and coding discipline at the C source level. 
The main properties are:

\textit{Source-level operation.}
\sname instruments the original firmware source code to inject its runtime library and inlined placeholders for activating its patching mechanism. 
There are two major factors that lead to this design choice. 
First, working at any level below source code would make it harder (if not impossible) to satisfy the ASIL-D classification requirements; for example, one requirement is to maintain bi-directional traceability between source code and object code~\cite{iso26262_traceability_requirements}, which would be disrupted when working at IR or binary level. 
Second, this allows us to be toolchain-agnostic, as different manufacturers may use different compilers for building their software, which makes working at other levels (especially the IR level) less feasible.
Additionally, it is also an automotive industry standard to work at the source code level when injecting code (i.e., source-to-source transformation), which also ensures that the modified code is still human-readable and can be easily reviewed. 

\textit{Deterministic behavior.}
\sname is crafted with deterministic and bound execution paths for its core mechanisms (placeholders, patch dispatcher, timing-buffer operations, "append-only" flash operations with recovery), so that its overhead is predictable. 
For example, we use a hash table to keep track of all active patches at runtime, which enables a deterministic fast lookup of patches. 
This also helps incorporate \sname into the system-wide WCET analysis, making it easier to integrate \sname into existing firmware development workflows. 
Our evaluations further confirm microsecond-scale stable performance overhead, and static memory usage that does not grow with the number of loaded patches. 

\textit{No hidden control/data flow, no unconditional jumps, no recursion.}
\sname does not introduce unstructured branches or recursion, and it prohibits patches from arbitrarily skipping subsets of instructions. 
The dispatcher function and patch code are invoked using normal calling conventions, and each patch can only either keep (execute) or skip an entire associated code section (with limited variants like skip-and-continue/skip-and-break for periodic sections), rather than jumping into the middle of a code section. 
This design avoids hidden flows and preserve explicit, analyzable control and data flows, though it is at the cost of some flexibility in patching.

\textit{No dynamic memory allocation.}
\sname uses only static memory; all data structures such as patch [metadata] table, runtime patch map, timing buffers, and other stateful data are allocated at compile time. 
This implies that configuration parameters such as the maximum number of tasks and patches must be determined at compile time, contributing to limited flexibility and relatively high but predictable memory usage. 
This design choice reflects ASIL-D guidelines that strongly discourage dynamic objects or heap allocation. 

\textit{Single-entry/single-exit (SESE) functions.}
All \sname related functions (e.g., dispatcher function, timing emulation function, patch storage function, etc.) are architected with precisely one entry and one exit point, following ISO 26262 recommendations for subprogram structure. 
This restriction simplifies control-flow analysis and supports formal or semi-formal verification. 

\textit{Restricted and explicit pointer usage.}
Pointer usage is intentionally cut down and confined to a few well-defined locations, primarily the dispatcher function and the patch calling convention, where pointers are used to provide the ability to overwrite the patched function's return value when needed. 
Furthermore, all type conversions are explicit, with no implicit casts, and the implementation follows defensive practices such as strong typing, consistent naming conventions, and initialization of variables. 

Finally, we expect developers to perform tests and analysis on their systems when integrating \sname according to the safety classification level their system is designed for. 
These tests and analysis shall include static code analysis, semi-formal verification (or even formal verification if possible), control and data flow analysis, and so on. 
As the logic of \sname is well-defined and self-contained, we do not expect that adding \sname would significantly increase the development burden of developers.

%% file: contents/patch_table.tex
\section{Task-Aware Patch Deployment}
\label{sec:patch_deployment}
\ardalan{Start each paragraph/section by telling us about the problem and motivating the rest of the write-up. This paragraph currently just starts by telling us what Patchlings does, not why it does it.}
\myles{Addressed here, but for some subsections, they're more of a "what" than a "why".}
To reduce the functional validation efforts needed for deploying a patch, \sname should bound the scope of a patch to only necessary places. 
Instead of overwriting the original firmware, \sname works by dynamically loading patches into a reserved memory section. 
More specifically, \sname puts placeholders across different locations, where they perform preliminary checks (e.g., if the current task has any patches) and then call into our patch dispatcher function. 
Once inside the dispatcher, if a patch is indeed found for the current context (determined by both the code location and the task), the dispatcher will execute the patch code.
This section discusses the design details of how patches are deployed in \sname. 


\subsection{Task-Awareness}
\label{sec:task_based_patching}
Similar to existing hotpatching approaches, \sname uses the LR register (as our patching mechanism resides in a standalone dispatcher function) to identify the current code location (i.e., the function being invoked), and subsequently determines if there is a patch available for it. 
However, a key distinction differentiates \sname is it operates with task-awareness. 
This means that \sname allows each patch to be triggered only when its target code location is being invoked by task(s) designated by the patch.
We made this design choice primarily for one reason: to reduce the validation efforts needed for deploying a patch.

A bug or vulnerability is often task-specific, meaning that it only manifests itself when the affected code is being executed in certain tasks.
However, by providing a patch to fix a bug or vulnerability in these problematic tasks, other tasks that also invoke the the same function would execute the patch code as well, which could results in changes in their behavior. 
As part of the safety requirements defined by ISO 26262, functional correctness of the entire system should always be guaranteed; therefore, such patching approach would require extensive verification and validation efforts to ensure that the patch does not introduce any unintended behavior or side effects in all tasks that invoke the same patched function.
By limiting the scope of a patch to only necessary task(s), we can significantly reduce such validation burden. 
Additionally, there are other benefits brought by task-aware patching, such as reducing logical complexity of patch development and restraining potential performance overhead.

With task-awareness in mind, we split the patch execution process into two steps: an inlined placeholder and a dispatcher function. 
The reason we separate the process into two steps is to reduce the performance overhead while limiting the storage overhead. 
We expect that in most cases, not all tasks would need patches; in fact, we think that it is highly likely that only a small subset of tasks have patches. 
Therefore, by putting a lightweight inlined placeholder that checks if there is any patch for the current task before calling into the dispatcher, we can avoid unnecessary calls into the dispatcher when there is no patch for the current task, reducing the overall system performance overhead.

\noindent
\begin{minipage}{\linewidth}
\begin{lstlisting}[language=C, style=VS2017, caption=C-style pseudo code of \sname placeholders for a general function.\myles{maybe move this to appendix if we really need to save space?}, label={lst:placeholder_4_function}]
  extern uint32_t patch_readiness;

  RETURN_TYPE funcA()
  {
      // Variable definitions
      ...
      RETURN_TYPE result = {0};
  
      // Patchlings runtime patch
      uint8_t cur_task_pid = getPID();
      uint8_t call_status = PATCHLINGS_CALL_STATUS_NORMAL;
      RETURN_TYPE call_result = {0};
      if (patch_readiness >> current_task & 1)
          call_status = patchDispatcher(cur_task_pid, (void *)&call_result);

      // Patchlings detour
      if (call_status == PATCHLINGS_CALL_STATUS_NORMAL)
      {
          // funcA code
          ...
      }

      // Patchlings result replacement
      if (call_status & HOT_PATCH_STATUS_OVERWRITE_RESULT_MASK)
          result = call_result;

      // Return statement
      return result;
  }
\end{lstlisting}
\end{minipage}

\subsection{Inlined Placeholder}
\label{sec:inlined_placeholder}
We use inlined placeholders to provide patching capability at various locations in the original firmware.
Our inlined placeholder mainly has three capabilities: checking if the current task has any patches (and calling the dispatcher if needed), skipping the original function's code (if instructed by a patch), and replacing the original function's return value (if instructed by a patch). 
There is a variant of our inlined placeholder that has a fourth capability of helping maintain timing behavior, which we will explain in Section~\ref{sec:timing_awareness}.

First, we assume there are a maximum of 32 patchable tasks running, though this number could be increased if needed (at the cost of performance).
The reason we set such limitation is to make sure a 32-bit global variable (called \texttt{patch\_readiness} in our system) can cover all tasks (where each bit represents a task), as the majority of microcontrollers are 32-bit based~\cite{Kamal_2015, microcontroller_market_analysis_2023}. 
The placeholder performs a simple check using \texttt{patch\_readiness}: examine if there are any patches for the current running task. 
This check is translated into only a few lines of instructions and can be performed without accessing any memory (in most cases). 

Second, we expect a patch to either be an addition of some code before the original function's code (e.g., filtering out some inputs or adding some logging) or a replacement of the associated code section (e.g., replacing the original logic). 
Hence, we provide the ability to keep or skip the execution of the original code (i.e., detouring), which is controlled by the patch code. 

Third, for general functions with non-void return types, we also provide the ability to overwrite the return value of the original function (i.e., result replacement), which is also controlled by the patch code. 

We now describe how placeholders are planted in different types of code locations, where each placeholder is associated with a fixed section of code. 
We classify all functions into two categories: general functions and RTOS task functions. 
A general function is a function that is invoked by other functions regularly, and a RTOS task function is a function that is directly invoked by the RTOS scheduler as a task. 
For general functions, we put the placeholder at the beginning of the function, as demonstrated in Listing~\ref{lst:placeholder_4_function}, and the placeholder is associated with the entire function body. 
For RTOS task functions, we expect them to have three phases: entry phase, periodic runtime phase, and exit phase. 
The entry phase is the code that is executed once when the task starts, the periodic runtime phase is the main loop of the task, and the exit phase is the code that is executed once when the task is about to terminate. 
Such structure is common in automotive systems, especially for those following the AUTOSAR framework~\cite{autosar_rtos_task_structure_mathworks, model_entry_point_functions_structure_mathworks}. 
We put a variant of the placeholder similar to the one for general functions at the beginning of each phase in a RTOS task function. 
In this way, we provide good flexibility for applying patches at different places, while having excellent coverage for patching and keeping both performance and storage overhead manageable. 
In both categories of functions, we provide the ability to keep or skip the execution of the original code, which is controlled by the patch code, though we offer two different skip modes for the runtime loop of RTOS task functions: skip-and-continue and skip-and-break, where the former means that the task will continue to execute the next iteration of the runtime loop, while the latter means that the task will exit the runtime loop and proceed to the exit phase. 
The other difference between the two categories is that there is a timing-safety preservation mechanism for RTOS task functions, which will be discussed in Section~\ref{sec:timing_awareness}. 
We also provide a concrete example of how placeholders are put in a RTOS task function in Appendix~\ref{app:placeholder_4_task}.


\subsection{Dispatcher}
Once a placeholder calls into the dispatcher, the dispatcher uses the LR register to determine if there is a patch available. 
For deterministic and fast patch lookup, we use a hash map to keep track of all existing patches at runtime, where the LR value serves as the unique identifier (key) for each patch, and it is used to identify the corresponding patch's metadata (i.e., \texttt{patch\_info}). 
As we are aware of the exact pattern of LR values, we build an optimized hash function to minimize collisions and assure deterministic lookup time. 
After finding a valid patch using the LR value, \sname checks if the current task is marked as active in that patch's \texttt{patch\_info}, which it would then execute the patch code. 
After executing the patch code, the dispatcher replaces the original function's result (if needed) and updates the active timing buffer (to be elaborated in Section~\ref{sec:timing_awareness}) for the current task before returning to the placeholder. 
Listing~\ref{lst:dispatcher} shows a C-style pseudo code of \sname's dispatcher. 

\noindent
\begin{minipage}{\linewidth}
\begin{lstlisting}[language=C, style=VS2017, caption=C-style pseudo code of \sname dispatcher.\myles{maybe put this in the appendix to save space?}, label={lst:dispatcher}]
  extern struct hashmap_t runtime_patch_map; // a hash table for managing patches at runtime
  extern uint32_t timing_buffers[32]; // assuming max 32 tasks

  uint8_t patch_dispatcher(const uint8_t pid, void* result) 
  {
      // Get LR register value
      uint32_t lr_value = __asm__reg_lr;

      // Look up a patch with the LR value
      struct patch_info_t* patch_info = hashmap_get(&runtime_patch_map, lr_value);

      uint8_t status = PATCHLINGS_CALL_STATUS_NORMAL;

      // Execute the patch if contextually appropriate
      if (patch_info != NULL && (patch_info->tasks & (1 << pid))) 
      {
          status = ((uint8_t (*)(void*))(patch_info->addr))(result);

          // Add the patch delay to the corresponding task's active timing buffer
          timing_buffers[pid] += patch_info->delay;
      } 

      return status;
  }
\end{lstlisting}
\end{minipage}

\begin{figure}
    \centering
    \includegraphics[width=0.75\columnwidth]{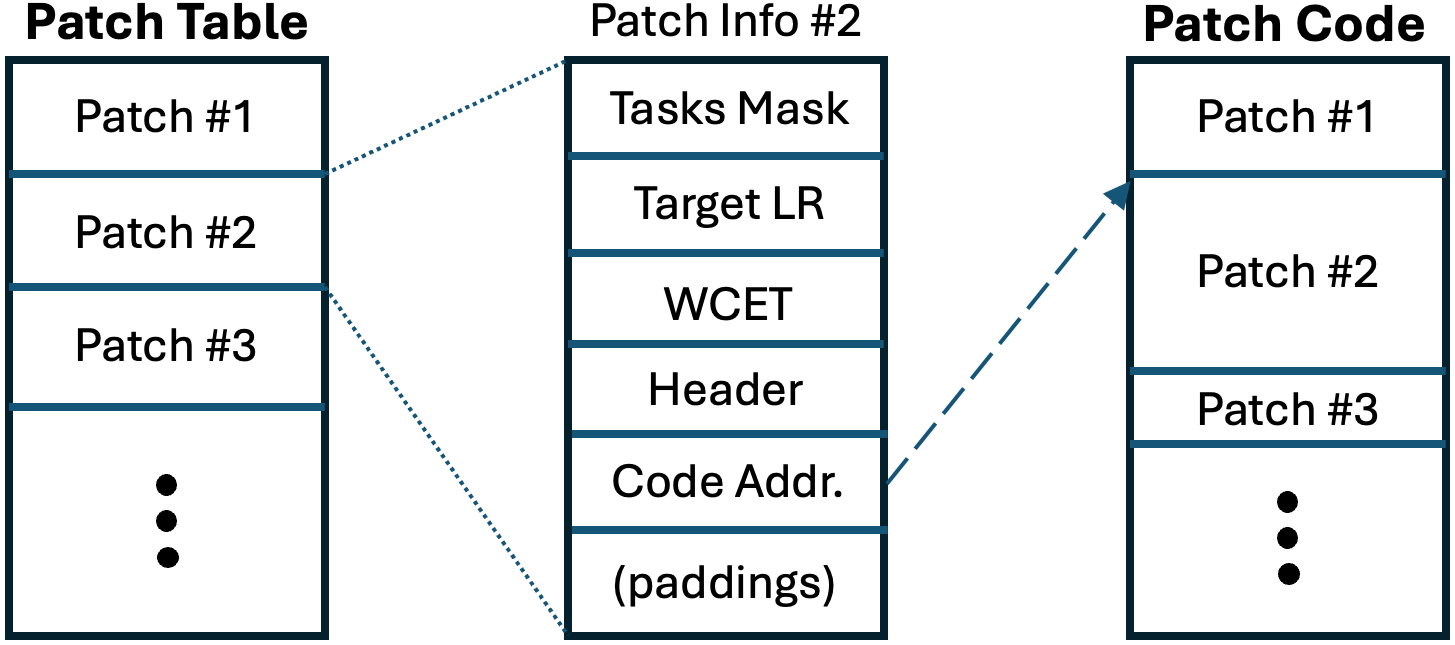}
    \caption{\em An example of how \texttt{patch\_info} is organized and used in the patch table.}
    \label{fig:patch_table}
\end{figure}

\subsection{Patch Table}
\sname uses a patch table to maintain all patches' \texttt{patch\_info}. 
Figure~\ref{fig:patch_table} shows how \texttt{patch\_info} is organized and used in the patch table. 
Each \texttt{patch\_info} records the following information: all related tasks, key identifier of the patch (i.e., expected LR register value), size of the patch, WCET of the patch (in $\mu$s), pointer to the patch code, and (optionally) some padding bytes for aligning the structure size to both the minimum write size of the underlying hardware and the sector size of the flash memory. 

%% file: contents/timing_aware.tex
\section{Timing-Aware Patch Execution}
\label{sec:timing_awareness}
In hard real-time systems, especially those automotive ones, timing correctness is as important as functional correctness. 
Whenever new code is added to such a system, developers usually need to conduct a complete set of static and dynamic timing evaluations, which is both time-consuming and notoriously error-prone. 
As we want to reduce the MTTM, requiring developers to perform such system-wide timing analysis after each patch is not a practical solution. 

Although our task-aware patch deployment routine bounds the scope of patching to designated tasks, it still requires developers to perform a complete system-wide timing re-evaluation after each patch. 
This is due to the fact that even there is only a small amount of timing behavior change in one patched task, it can still affect the overall system timing behavior in a complex way, as it may interact with other tasks through various shared resources (e.g., shared memory, I/O devices, etc.) or even send signals outside of the system.

To solve this challenge, \sname seeks to preserve the timing behavior of the original system by providing timing-aware patching. 
More specifically, \sname integrates a timing buffer to the overall system timing budget to offset the timing impact brought by new patch code. 
This allows \sname to guarantee that as long as (1) measured WCETs of tasks and patches are not violated (2) and our timing buffer is not exhausted, the upper bound on CPU time consumed by tasks will not exceed their respective deadlines, or in other words, the deadlines of all tasks are preserved. 


\subsection{Task-Associated Timing Buffer}
When designing \sname, our first idea was to put an extra timing buffer for each inlined placeholder, where the buffer size is equal to a certain percentage of the original function's WCET. 
The timing buffer is used to emulate the timing overhead of the patch code even when there is no patch applied. 
In this way, whenever a patch is applied to that function, as long as the patch code's WCET is smaller than the buffer size, by deducting the WCET of the patch code from the timing buffer, the overall timing behavior of that function remains unchanged.
However, after further analysis, we found that this naive design has two critical problems.
First, a function's execution time (especially its WCET) can vary significantly, as it can be called by different tasks in different contexts (e.g., different input parameters, different system states, etc.), and this could directly affect the amount of workload of the function. 
Second, even if the WCET is representative enough for that function, coming up with a proper buffer size from a function level WCET is still very challenging. 
A small buffer size could be insufficient to accommodate the WCET of the patch code, especially when the patch code introduces new logic; a large buffer size, on the other hand, might waste too much timing resource and disturb the overall system performance. 

\begin{figure}
    \centering
    \includegraphics[width=1\columnwidth]{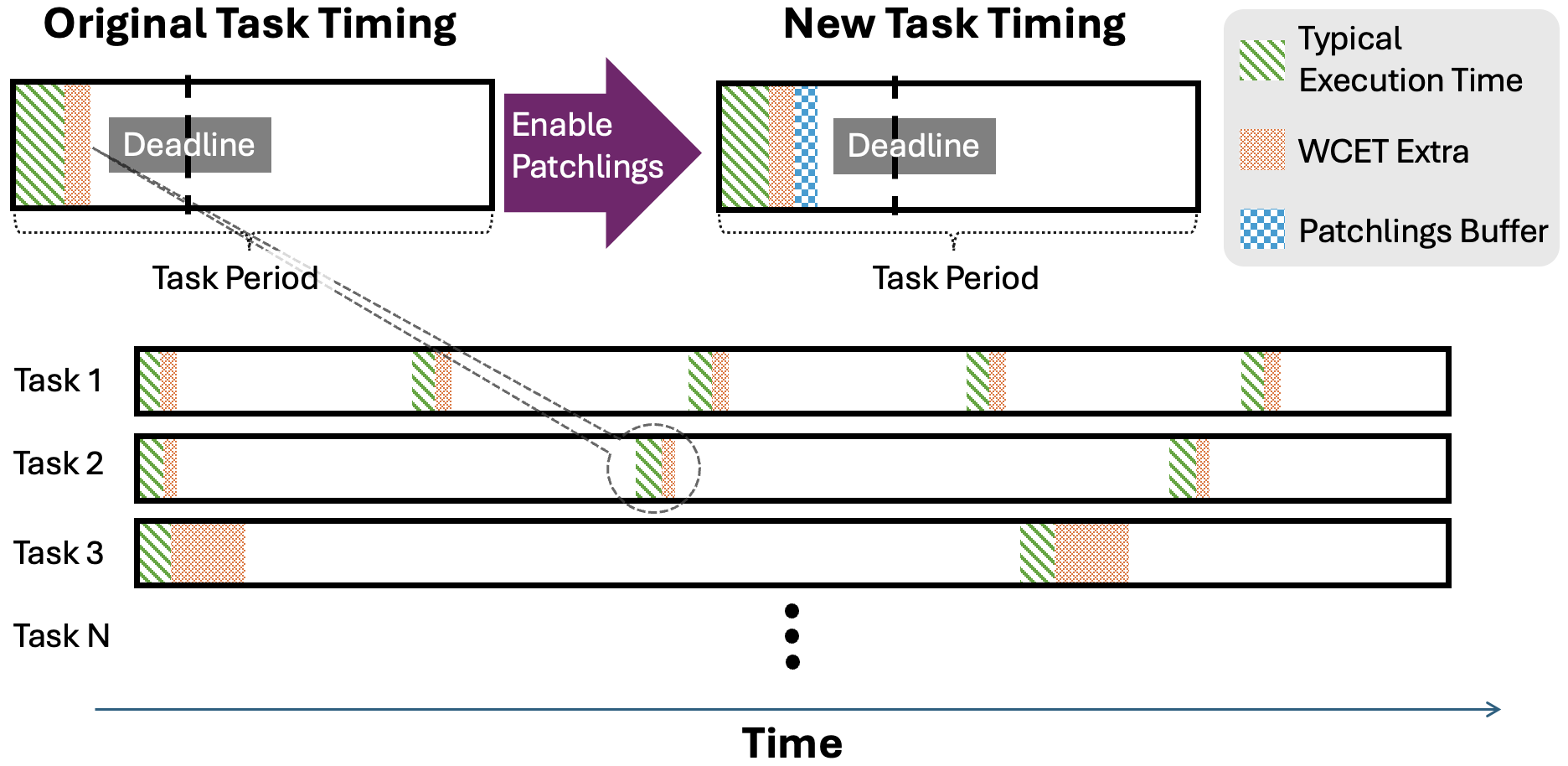}
    \caption{\em Illustration of how \sname timing buffer is integrated with timing budgets of tasks.}
    \label{fig:tasks_timing_buffer}
\end{figure}

To address these problems, we look into how timing behavior is usually characterized in automotive systems. 
We found that periodic tasks are treated as the basic unit for timing analysis and verification. 
Based on this observation, we design the timing buffer mechanism in \sname to work at the task level instead of the function level. 
As shown in Figure~\ref{fig:tasks_timing_buffer}, we add a timing buffer to each periodic task based on the WCET of its periodic code. 
In this way, all functions invoked within a task share the same timing buffer, which not only simplifies timing analysis and verification but also allows a more flexible allocation of timing resources among patches. 
It is noteworthy mentioning that timing buffers are meant to stay static, no matter how the task's execution time varies at runtime.

There are two special considerations regarding our task-associated timing buffer. 
First, other than periodic code being executed at regular intervals, there is also sporadic code (e.g., boot-time initialization code, one-time entry/exit code for periodic tasks, etc.), which could require patching capability as well. 
In such cases, we need to allocate additional timing buffers for sporadic code. 
Second, by default, when emulating the consumption of a timing buffer, we make the CPU execute a busy-wait loop (i.e., keep executing NOP instructions) to consume the extra time. 
We do not use sleep related functions, as they may introduce additional variability in the overall system timing behavior, though developers can choose to overwrite our design if needed. 

\begin{figure} 
    \centering
    \includegraphics[width=1\columnwidth]{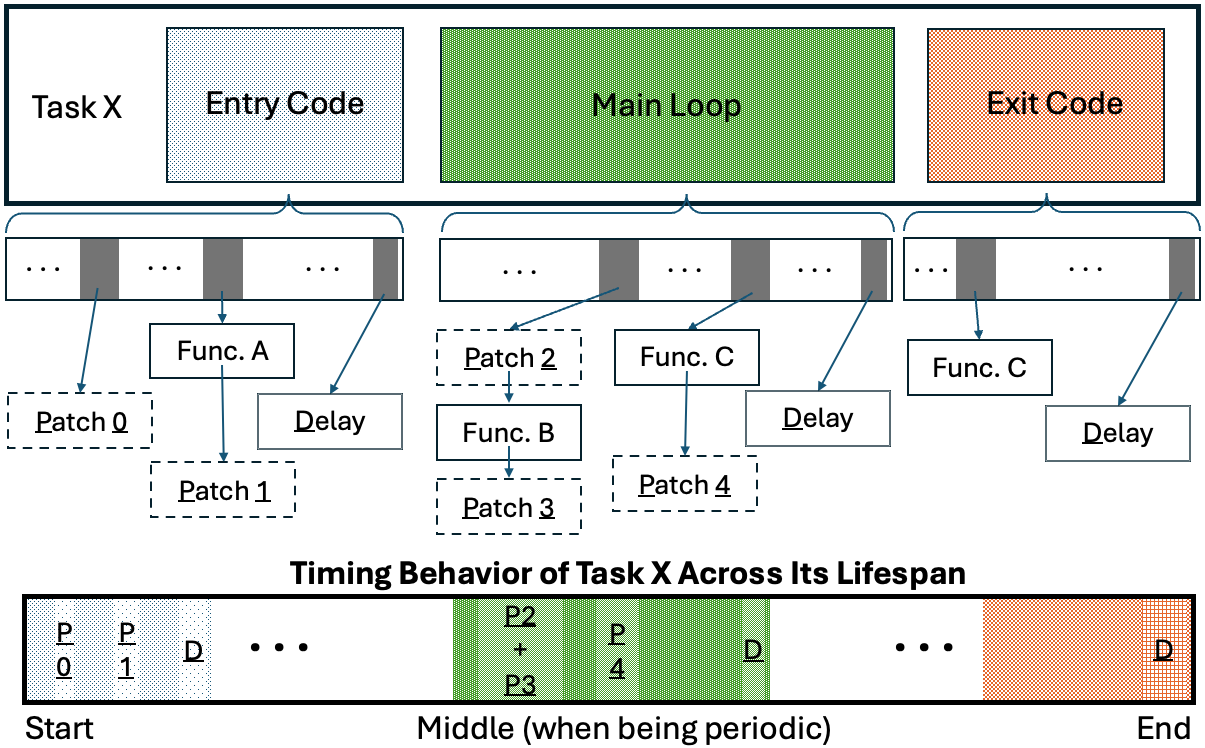}
    \caption{\em A concrete example showing how \sname timing buffer(s) are integrated into the system.}
    \label{fig:timing_example}
\end{figure}

\subsection{Patch Execution}
We offer timing buffer for both periodic and sporadic code. 
At runtime, we keep track of the active timing buffer for each task, which records the remaining time that can be used for execution of patches in that task. 
Whenever we enter a new timing buffer protected code section (either periodic or sporadic), we first copy the predefined timing buffer size into the active timing buffer for current task. 
Within our dispatcher function (as shown in Listing~\ref{lst:dispatcher}), after executing the patch code, we update the active timing buffer for the corresponding task by deducting the WCET of the patch code from the remaining time buffer.  
At the end of that code section, we run our delay function to consume the remaining time in the current task's timing buffer. 
Figure~\ref{fig:timing_example} demonstrates a concrete example of how the timing buffer is utilized throughout the lifetime of a task that contains both periodic and sporadic code.

\subsection{Integration with Development Workflow}
Our buffer-based timing safety-preserving mechanism is built into the placeholder code. 
This means that it can be automatically added along with other placeholder code during the instrumentation process, which is done by leveraging LLVM Clang's LibTooling. 
However, as it is a timing-related mechanism, and the mechanism itself introduces changes to the timing behavior, enabling it requires a 2-step process. 
First, during the instrumentation process, the timing buffer code (with each timing buffer size set to a default dirty value, which is also defined as volatile to prevent any compiler optimizations) is added along with the placeholder (for both RTOS task functions and sporadic functions like \texttt{main()}). 
Second, using the (intermediate) compiled binary from the first step, developers can use various WCET analysis tools~\cite{es_prj_wcet_tool, ballabriga2010otawa, hardy2017heptane, holsti2002bound_t} to determine the WCETs, and then update the timing buffer sizes accordingly. 
We opt to not integrate a WCET analysis tool into our instrumentation process, because different automotive ECUs, software architectures, and manufacturers may have different approaches (with some of them being proprietary) to perform WCET analysis. 
Furthermore, to our knowledge, the automotive industry prefers to use hardware-in-the-loop WCET analysis, which is only achievable with such 2-step process. 
Finally, after the 2-step process, the system will be ready for deployment. 

As for patch development, after a patch is developed and compiled, developers should measure and put the WCET of the patch code into the patch's \texttt{patch\_info}. 
In case the patch is meant to replace the original function code (i.e., the original function code will be skipped), the original function's WCET will also need to be analyzed and deducted from the patch's WCET to get the actual timing behavior after the patch is applied. 
Another consideration is that if the patch code for replacement has a smaller Best Case Execution Time (BCET) than the original function code, we expect the developers to either update the WCET in the patch's metadata or introduce additional delay within the patch code to compensate for the timing difference. 
This is because, for example, if a patch makes a function finish much earlier than the original function, it may as well alter the system's timing behavior in an undesirable way, potentially causing other tasks or even other systems to behave unexpectedly. 

We do acknowledge that integrating \sname's timing safety-preserving mechanism into the system requires additional effort from developers. 
However, we argue that the overall development effort is already drastically reduced compared to other update mechanisms, which all require a complete system-wide validation and verification process for timing safety. 
Moreover, we also believe that by providing a more structured approach to preserve timing safety, we can lower the risk of introducing timing-related bugs into the system due to human errors. 


%% file: contents/persistency.tex
\section{Flash-Based Patching}
\label{sec:patch_persistency}

In order to fulfill our vision of future software updates for automotive ECUs, patches in the hotpatching system must survive system resets and integrate with the XIP architecture common in automotive ECUs, where code runs directly from flash memory.
Therefore, \sname implements a lightweight "append-only" flash-based storage system designed for determinism and real-time safety. 
To the best of our knowledge, none of the existing general-purpose embedded file systems~\cite{zephyr_vfs, LittleFS, FatFs, reliance-edge, tuxera-edgefs-cert} can provide the level of predictability and robustness that \sname requires. 

Although \sname can also operate in RAM-based storage mode, for the rest of this paper, we will focus on flash-based storage mode.
We assume a flash region is reserved for \sname, where the patch [metadata] table is put at the beginning of the region, followed by all the patch code. 
Additionally, we reserve a sector towards the end of the region to act as a recovery area.

As for security of patches, \sname does not introduce a new trust model. 
It relies on existing secure boot, update authentication, and authorization infrastructure to control write access to patch flash regions and redirection tables. 
Further security considerations are discussed in Appendix~\ref{app:security_considerations}.



\subsection{Append-Only Patch Loading}
\label{sec:append_only_patch_loading}

We assume that the flash memory is completely erased (i.e., all bits are 1) before \sname is first used, which is typically the case for NOR flash used in automotive ECUs. 
Our core storage strategy is "append-only."
When a new patch is loaded, its binary code and metadata (\texttt{patch\_info}) are written sequentially to the end of their respective storage areas in flash. 
Crucially, existing patches are never modified in place. 
This design provides two key benefits.
First, safety of the existing system is guaranteed, as we eliminate the risk of corrupting existing patches. 
Second, the hard real-time property of the system is preserved, as we avoid the need for slow, blocking flash erasure during normal operation; we analyze the hard real-time property of \sname in Section~\ref{sec:real_timeness_evaluation}.



The patch loading process is a simple two-step sequence: first, the patch's binary code is written to flash, followed by its \texttt{patch\_info}. 
This order, combined with the atomicity of flash writes, guarantees that any valid \texttt{patch\_info} will always point to a complete patch code. 
Figure~\ref{fig:patch_loading} (Appendix~\ref{app:visualizing_patch_loading}) visualizes the process of how \sname writes a patch to flash. 

\begin{figure} 
    \centering
    \includegraphics[width=1\columnwidth]{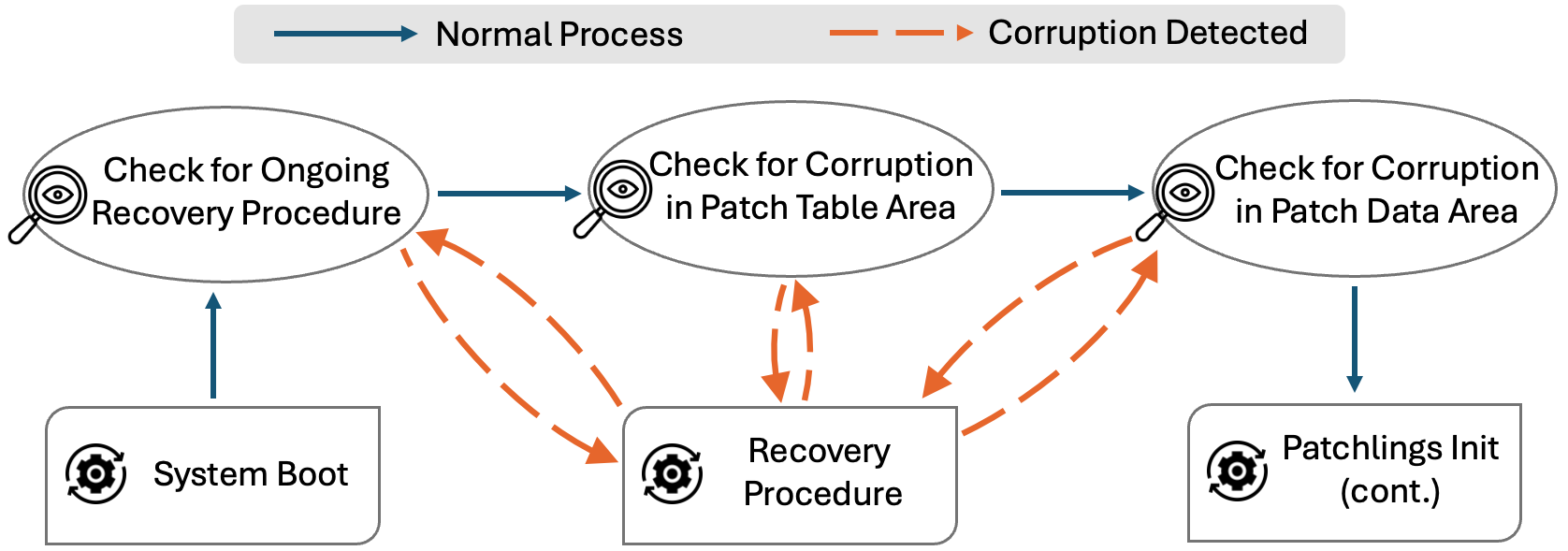}
    \caption{\em Flow of how \sname checks and handles corruption during boot-time initialization.}
    \label{fig:flash_patch_table_init}
\end{figure}

\subsection{Corruption Resilience and Recovery}
\label{sec:corruption_resilience_and_recovery}
While the "append-only" design is robust, power loss during a write can still corrupt the single flash sector being written to. 
\sname is designed to confine the impact and recover gracefully. 
More specifically, \sname offers the ability to recover from any dirty (hardware corruption) or incomplete (software corruption) write to the patch table or patch code area. 

At boot time, the system performs a series of integrity checks. 
It scans both the patch table and the patch code area in flash for any signs of an incomplete write or corrupted data. 
If a corrupted sector is detected, a recovery procedure is initiated. 
This procedure uses the dedicated recovery sector in flash as temporary storage to safely restore the last known good state of the affected sector, ensuring the integrity of the patch storage before the main system code is executed. 
This entire mechanism assures that patch persistence does not compromise the safety and robustness of the system. 

In addition, we also assume the possibility of power loss during the recovery procedure. 
In this case, \sname will try to resume the recovery procedure from where it left off when the power is restored. 

Figure~\ref{fig:flash_patch_table_init} illustrates the flow of how \sname checks and handles corruption during boot-time initialization. 
We also provide a detailed technical breakdown of the recovery procedure in Appendix~\ref{app:recovery_procedure}.

%% file: contents/prototype_and_evaluation.tex
\section{System Implementation}
\label{sec:implementation}
There are two main components in our prototype: instrumentation toolchain and runtime library. 
The total development effort includes approximately 3500 lines of C++ and Python for the instrumentation toolchain, 4000 lines of C code for the runtime library, 700 lines of helper scripts (e.g., bash, Python, etc.), and 1200 lines of C code for evaluation purposes (e.g., CAN-based patch sender, CVE patches, etc.).

Our prototype is developed on an automotive-grade hardware platform (i.e., \textit{NXP S32K148EVB-Q176}~\cite{nxp_s32k148_evb}), which features a single-core ARM Cortex-M4F processor, 256 KB of SRAM, and 2048 KB of flash memory. 
This represents a common class of hardware used in modern automotive ECUs. 
To demonstrate portability, we integrated the \sname runtime library with two widely used RTOSes: FreeRTOS and Zephyr OS. 

\subsection{Instrumentation Toolchain}
The first key component of our implementation is a toolchain that instruments the target firmware's source code. 
This process prepares the firmware to be patchable by \sname. 
We leverage LLVM Clang's LibTooling~\cite{llvm_libtooling}, which provides APIs for parsing and modifying firmware source code using Clang's frontend and the Abstract Syntax Tree (AST), to automatically identify function types and put placeholders in the correct locations (as described in Section~\ref{sec:inlined_placeholder}).
Using the instrumented firmware source code, we can then build an intermediate firmware binary for the 2-step timing buffer integration, where we use an open-source static WCET analysis tool~\cite{es_prj_wcet_tool} to measure WCETs and configure our timing buffers. 
Finally, we put correct timing values into the firmware source code and then build the final firmware binary. 

\subsection{Runtime Library}
The second component is the \sname runtime library, which is a self-contained and highly-portable library that manages the loading and execution of patches at runtime. 
At boot time, the library's initialization routine scans the flash memory, performs integrity checks, and populates the hash-table-based runtime patch map. 
The \sname task runs at the lowest priority, listening for new patches delivered over a communication bus (we use Controller Area Network (CAN) for our prototype, but any bus is supported). 
When a patch is received, the task first temporarily stores it in RAM, and then loads it into flash memory (using the "append-only" mechanism described in Section~\ref{sec:append_only_patch_loading}). 
Finally, the patch becomes active and is triggered at its corresponding context (using the task-based mechanism described in Section~\ref{sec:patch_deployment}). 

\section{Evaluation}
\label{sec:evaluation}
We evaluate our prototype to measure its system overhead, its impact to the hard real-time property of the system, and its effectiveness in patching real-world CVEs. 
All experiments are conducted on our \textit{NXP S32K148} evaluation board, as described in Section~\ref{sec:implementation}. 
Unless otherwise specified, we use the following configuration:

\begin{itemize}[noitemsep, nolistsep]
    \item FreeRTOS v11.2.0 or Zephyr v3.5.0
    \item GNU GCC 6.3.1 for FreeRTOS and GNU GCC 12.2.0 for Zephyr (both with \textit{-O2} optimization)
    \item 48 MHz on FreeRTOS and 80 MHz on Zephyr (CPU clock)
\end{itemize}

We acknowledge this is a proof-of-concept evaluation, and performance may differ in production environments with varying hardware and software configurations. 

\subsection{System Overhead}
\label{sec:system_overhead}
\textbf{Storage.}
In FreeRTOS, \sname adds about 8.19 KB of binary size, which is a 6.45\% increase over the baseline FreeRTOS; in Zephyr, it adds about 61.71 KB of binary size, which is a 6.34\% increase over the baseline Zephyr. 
The larger absolute increases on Zephyr is attributable to its more complex build system and the preliminary nature of its board support package (where we had to port some code from FreeRTOS to Zephyr, e.g., the flash driver). 
The \sname placeholder itself is lightweight, adding about 8 to 10 lines of C code per instrumented location, which in our evaluation environment results in about 29 to 35 lines of assembly instructions in FreeRTOS and about 20 to 24 lines of assembly instructions in Zephyr. 
If further translated to size in bytes, it occupies about 68 to 80 bytes of storage in FreeRTOS and about 46 to 60 bytes of storage in Zephyr. 

\begin{figure}
    \centering
    \includegraphics[width=0.48\columnwidth]{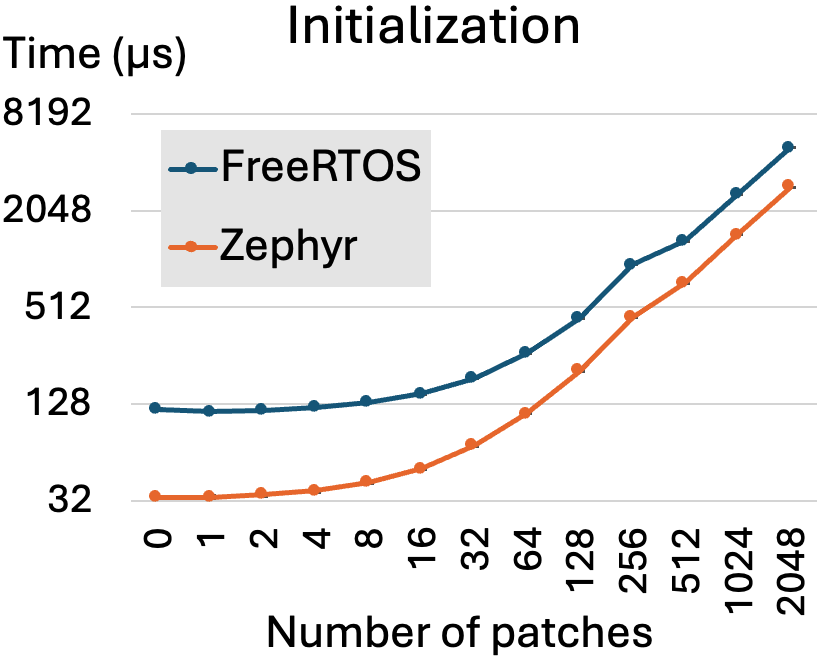}
    \includegraphics[width=0.48\columnwidth]{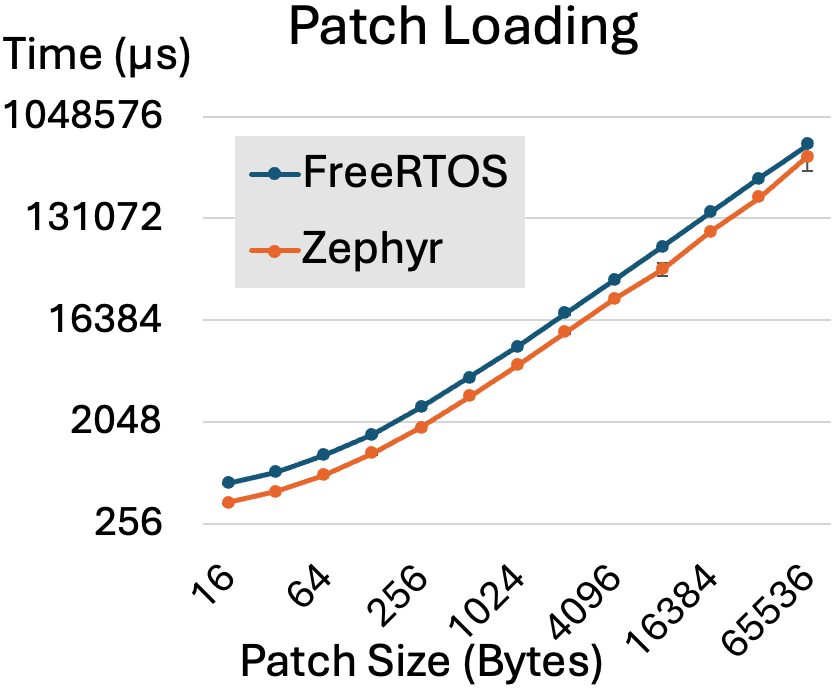}
    \caption{\em Left: performance of initialization (with different number of patches stored) at boot time. Right: performance of loading a single patch (with different sizes).\ardalan{Some of the texts in the figures are too small.}}
    \label{fig:eval_init_and_patch_loading}
\end{figure}

\begin{figure}
    \centering
    \includegraphics[width=0.48\columnwidth]{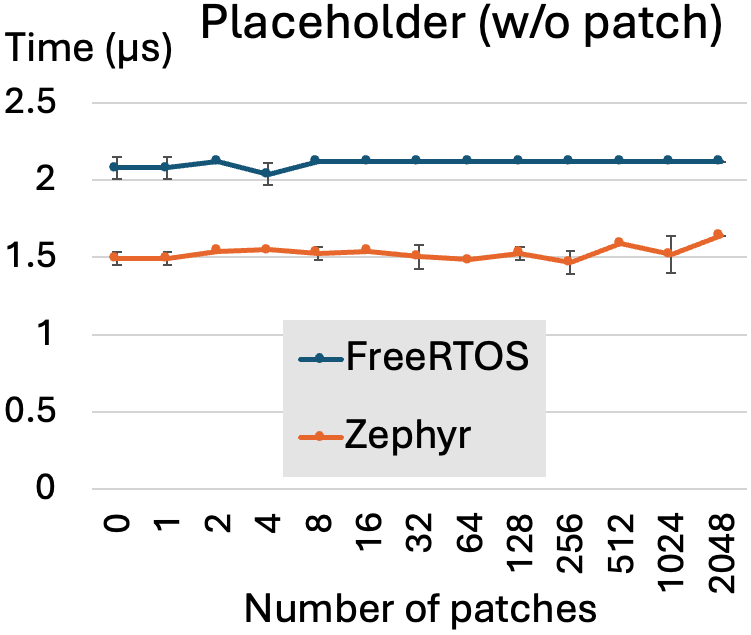}
    \includegraphics[width=0.48\columnwidth]{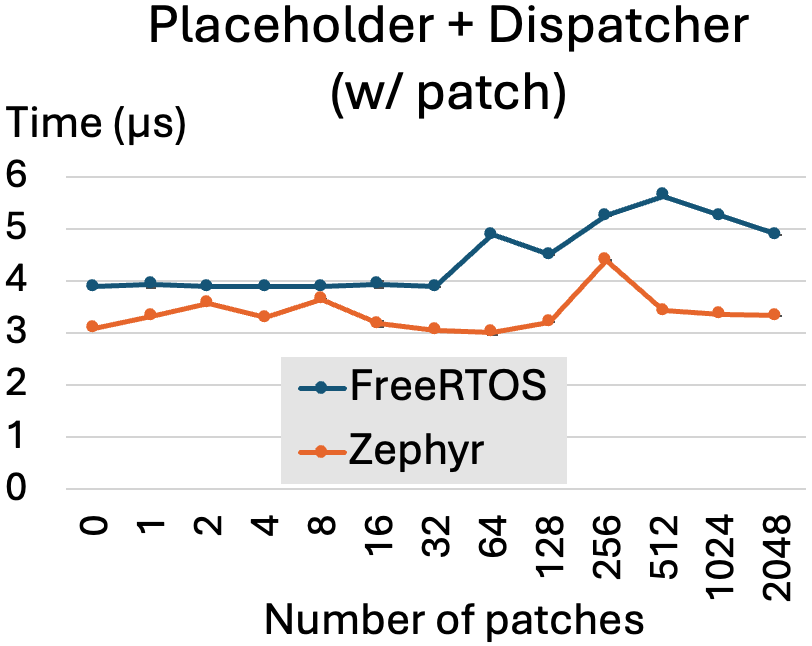}
    \caption{\em Left: performance of \sname placeholder (with no patch available for the current task). Right: performance of \sname placeholder and dispatcher (with patches available for the current task).}
    \label{fig:eval_placeholder_and_dispatcher}
\end{figure}

\begin{figure}
    \centering
    \includegraphics[width=0.48\columnwidth]{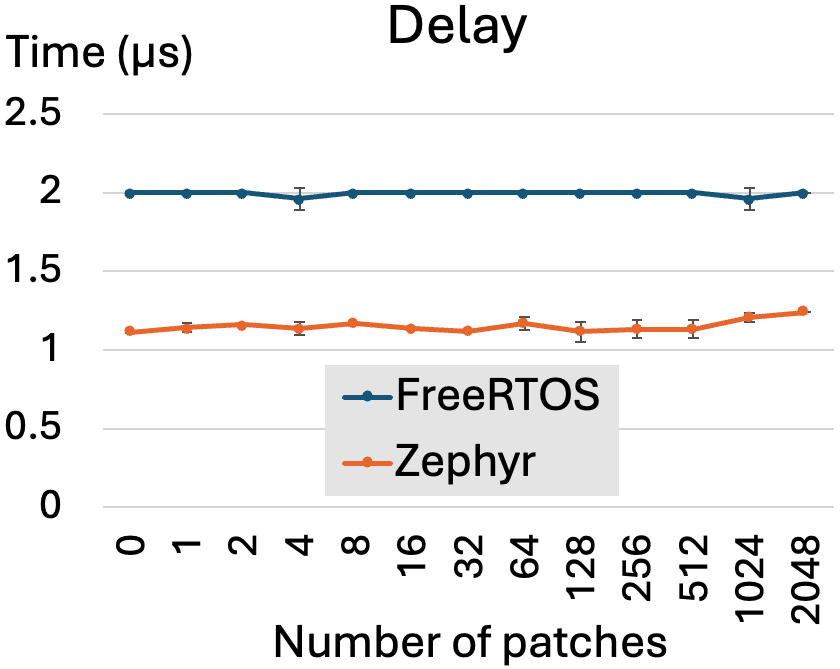}
    \includegraphics[width=0.48\columnwidth]{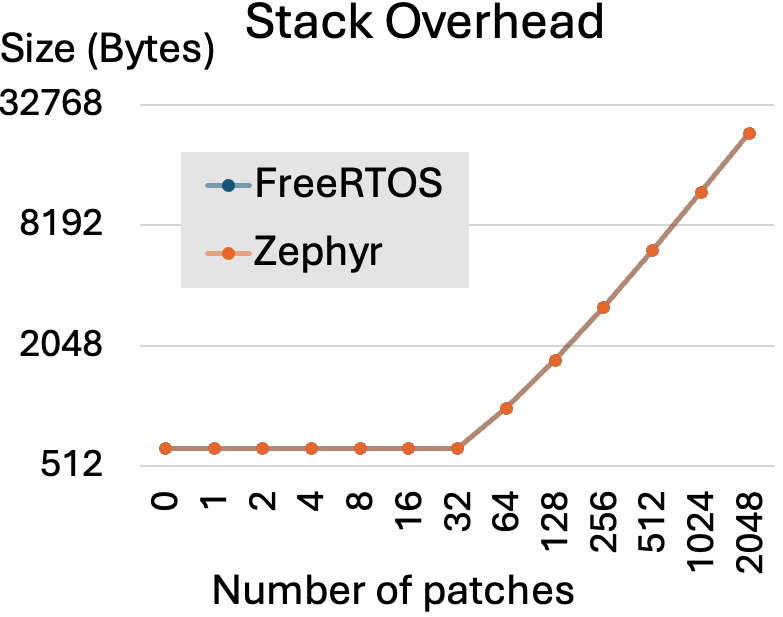}
    \caption{\em Left: performance of \sname timing buffer delay mechanism (without any timing buffer). Right: extra stack usage of \sname (without any burst usage).}
    \label{fig:eval_delay_and_stack_overhead}
\end{figure}

\textbf{Performance.}
The runtime overhead of \sname is deterministic and consistently low. 
As shown in Figure~\ref{fig:eval_init_and_patch_loading}(left), the overhead of \sname initialization scales linearly with the number of stored patches. 
Similarly, whenever a new patch becomes ready (i.e., in our case is downloaded to local SRAM by our CAN receiver code), as displayed in Figure~\ref{fig:eval_init_and_patch_loading}(right), the overhead of registering the patch (including saving the patch to flash) scales linearly with the size of the patch.  
When the current task has no patch available, any \sname placeholder triggered within it adds an overhead of about 2.1 $\mu$s in FreeRTOS and 1.5 $\mu$s in Zephyr, as shown in Figure~\ref{fig:eval_placeholder_and_dispatcher}(left). 
When the current task has patches available, any \sname placeholder triggered within it will invoke the dispatcher function as well, and in this case, the overhead becomes around 3.9 $\mu$s in FreeRTOS and 3-3.3 $\mu$s in Zephyr for up to 32 patches, as shown in Figure~\ref{fig:eval_placeholder_and_dispatcher}(right). 
When the maximum number of patches goes beyond 32, the overhead of the placeholder and dispatcher function increases slightly, but it is still very small. 
The delay mechanism (i.e., a function for consuming the current tasks's remaining timing buffer) also adds an overhead of about 2 $\mu$s in FreeRTOS and 1.2 $\mu$s in Zephyr, as shown in Figure~\ref{fig:eval_delay_and_stack_overhead}(left). 

\textbf{Memory.}
Figure~\ref{fig:eval_delay_and_stack_overhead}(right) shows the extra stack usage of \sname, which is about 630 bytes in both FreeRTOS and Zephyr when the maximum number of patches is 32. 
The major contributor to the stack usage is our runtime patch map, which is implemented using a hash table with a load factor of 0.7. 
\myles{Note that I also have an alternative implementation (already evaluated) where I replaced the hash table with a binary search tree, though I don't think it would help us to make any claims here and I decided to not talk about it.}
There is also a burst usage of stack when a patch is being loaded, where we currently use SRAM as the temporary storage for the patch, and the size of the burst usage is determined by the maximum allowed size of a patch. 
This can potentially be optimized by not waiting all the patch data to be received before moving them to flash memory. 

\input{contents/cves_table.tex}

\subsection{Impact on Real-Time Property}
\label{sec:real_timeness_evaluation}

A primary mission of \sname is to enable hotpatching without compromising the hard real-time property of RTOS. 
To validate this, we create a "brake-by-wire" case study. 
The system include a high-priority, periodic "brake" task with a strict 500 $\mu$s deadline~\cite{Todeschini2015, Iyenghar2020}, which is responsible for controlling the brake system of a car. 
We then deploy a patch to an unrelated, lower-priority task. 

In a conventional system, such a patch would mandate a full system-wide timing re-validation. 
Our results show that with \sname, the high-priority "brake" task never misses its deadline. 
Thanks to our "append-only" storage strategy, we manage to keep the blocking interval of flash related operations very small (only about 68.66 $\pm$ 0.08 $\mu$s in FreeRTOS and 44.04 $\pm$ 0.1 $\mu$s in Zephyr). 
This would be impossible if we were to use a conventional flash memory management strategy, where we would need to perform a flash erasure, which needs a significant amount of blocking time (about 5920 $\pm$ 10 $\mu$s in both FreeRTOS and Zephyr). 

There is one special case, where the device experiences a power loss during the patch loading process and results in corrupted flash memory. 
In this case, \sname needs to execute the recovery routine at boot time to restore the flash memory to the last known good state. 
During the recovery process, \sname performs flash erasures.
Although this means that the recovery process is lengthy (about 20000 $\mu$s to 25000 $\mu$s in both FreeRTOS and Zephyr), it can still be interrupted by other time-sensitive operations.
This is because the flash erasure itself can be suspended (and resumed at a later time), where the suspension operation takes about 58.95 $\pm$ 0.51 $\mu$s in both FreeRTOS and Zephyr. 
Consequently, an emergency task may potentially be executed during the flash erasure, with only a very short delay. 
We do want to note that in order to have a flash erasure to be completed successfully, there must be a few rounds of erasing intervals that are at least 2458.95 $\pm$ 0.51 $\mu$s in both FreeRTOS and Zephyr, as it takes time for the flash controller to make meaningful progress on erasing a flash sector. 
There are also workarounds for this special case. 
For example, code that need to be executed during the flash erasure can be moved to a different flash bank (that is not utilized by \sname); alternatively, for devices with only a single flash bank, the code can also be executed from SRAM directly. 
Furthermore, the recovery routine is only triggered at boot time, where we expect critical tasks are not yet initialized. 

\subsection{Addressing Real-World CVEs}
To demonstrate real-world applicability, we use \sname to patch a set of CVEs affecting FreeRTOS and Zephyr. 
For each CVE, we reproduce the vulnerability and develop a patch based on the official fix. 

As shown in Table~\ref{tab:cves}, \sname successfully applies a patch to mitigate each vulnerability. 
Our results also show that patch performance is highly dependent on the patch's design. 
The patch for CVE-2020-10067, for example, has a significant performance impact (about 3x slowdown), likely due to poor cache locality when accessing original firmware data and functions. 
This confirms that while \sname provides a robust mechanism for deploying patches, developers must still consider the performance implications of their patch code.

%% file: contents/cves_table.tex
\begin{table*}[t]
\centering
\resizebox{0.9\textwidth}{!}{%
\begin{tabular}{l l l r r r r r}
\toprule
\textbf{CVE No.} & 
\textbf{Type} & 
\textbf{OS} & 
\makecell[r]{\textbf{Hotpatch}\\\textbf{Size}\\\textbf{(Bytes)}} & 
\makecell[r]{\textbf{Native Fix}\\\textbf{Exec. Time}\\\textbf{($\mu$s)}} & 
\makecell[r]{\textbf{Hotpatch Fix}\\\textbf{Exec. Time}\\\textbf{($\mu$s)}} & 
\makecell[r]{\textbf{Execution}\\\textbf{Overhead}} & 
\makecell[r]{\textbf{Memory}\\\textbf{Overhead}\\\textbf{(Bytes)}} \\ 
\midrule

2021-32020 & Bound Checking       & FreeRTOS  & 172 & $7.16 \pm 0.08$   & $11.5 \pm 0.0$    & 60.54\%  & 24 \\
2021-31571 & Integer Overflow     & FreeRTOS  & 48  & $11.5 \pm 0.0$    & $15.5 \pm 0.0$    & 34.78\%  & 16 \\
2021-31572 & Integer Overflow     & FreeRTOS  & 16  & $13.88 \pm 0.0$   & $17.38 \pm 0.0$   & 25.22\%  & 16 \\
2020-10023 & Buffer Overflow      & Zephyr OS & 120 & $12.6 \pm 0.0$    & $16.69 \pm 0.0$   & 32.46\%  & 32 \\
2020-10024 & Privilege Escalation & Zephyr OS & 40  & $200.63 \pm 1.11$ & $207.57 \pm 0.66$ & 3.46\%   & 20 \\
2020-10067 & Integer Overflow     & Zephyr OS & 72  & $1.55 \pm 0.0$    & $6.01 \pm 0.0$    & 287.75\% & 16 \\
2023-3725  & Stack Overflow       & Zephyr OS & 88  & $3.67 \pm 0.0$    & $7.32 \pm 0.0$    & 99.46\%  & 32 \\ 
\bottomrule
\end{tabular}%
}
\caption{\sname patches for different CVEs and their overheads. Note that the memory overhead in this table refers to the extra memory footprint brought by the patch, not including the overall \sname memory overhead.}
\label{tab:cves}
\end{table*}

%% file: contents/discussions.tex
\section{Discussions}

While \sname provides a robust framework for hotpatching automotive systems, there are several important considerations and areas for future work. 
We discuss the topics of patch revocability, patch development effort and automation, strategic placement of placeholders, and the generalizability of \sname on other mission-critical real-time domains.

\subsection{Revocability}
In its current design, \sname does not support direct patch revocation. 
Although a simple "revocation" can be achieved by deploying a new, empty patch with the same identifier to overwrite the old one, implementing a full-featured revocation system introduces significant challenges.

Properly revoking a patch requires careful consideration of the system's state. 
For instance, if a patch is revoked, should the system attempt to revert data structures or variables modified by that patch?
How should the system handle a patch that is revoked while being executed? 
Furthermore, from a persistence standpoint, revoked patches would either need to be removed from flash storage, which introduces complex garbage collection and potential fragmentation issues, or be marked as "revoked," which consumes storage space and may also lead to non-deterministic overall system state.
Given these complexities, we leave full support of patch revocation as future work.


\subsection{Patch Development Effort}

\sname is designed for mission-critical systems where the safety of patches is paramount. 
Consequently, it prioritizes a verifiable and deliberate patch development process over full automation. 
We expect that patches will be carefully designed, developed, and subjected to rigorous [but self-contained] testing rather than being generated automatically from code commits.  


However, we believe that verifiable automation can still play a valuable role. 
Techniques from research like AutoPatch~\cite{salehi2024autopatch} could be adapted to assist in the patch creation process.
Furthermore, recent advances in using Large Language Models (LLMs) to fix code vulnerabilities~\cite{llm_vulnerable_code_localization_0, llm_vulnerable_code_localization_1} could also be explored to generate initial patch candidates, which would then be subjected to strict verification. 
This semi-automated approach could accelerate development while maintaining the high safety standards required in the automotive domain. 

\subsection{Placeholder Placement}
\label{sec:placeholder_placements}

The effectiveness and efficiency of \sname relies on the strategic placement of placeholders. 
Our current approach, which uses LLVM/Clang LibTooling to automatically instrument bodies of each general function and the entry/periodic/exit sections of each task function, provides broad coverage. 
However, this may introduce unnecessary overhead if only a few functions are likely to be patched. 


To optimize this, developers can manually inject placeholders into specific, high-risk code regions, though this requires deeper system knowledge and modeling/planning effort. 
A promising direction for future work is to develop automated techniques for optimal placeholder placement. 
For example, static analysis could identify critical code paths or functions with a history of vulnerabilities. 
Alternatively, machine learning models could be trained to analyze source code and suggest placements based on patterns learned from existing codebases, balancing patching flexibility and overhead. 

\subsection{Generalizability}
We design \sname specifically around automotive systems, but many of the benefits brought by \sname can be generalized to other types of real-time embedded systems as well, especially those that have hard real-time requirements and safety constraints, such as industrial control systems, medical devices, aerospace systems, and so on.
For example, in implantable medical devices (IMDs) like pacemakers, it is crucial to ensure that software updates do not compromise the device's real-time performance and safety. 
One of the key reasons that hotpatching is not yet adopted in these systems is the concern of unpredictable functional and timing behavior after applying patches, which could land the device in an undeterministic state. 
The goal of \sname matches well with the needs of these systems, as it aims to provide a reliable and predictable hotpatching solution that can ensure the safety and real-time performance of the system even after applying patches. 
We believe that \sname opens up new possibilities for adopting hotpatching in these mission-critical real-time embedded systems beyond automotive.



%% file: contents/relatedwork.tex
\section{Related Work} 
\label{sec:related_work}

\begin{table}[t]
  \centering
  \setlength{\tabcolsep}{3pt}
  \footnotesize
  \begin{threeparttable}
    \begin{tabular}{lrrrr}
      \toprule
      \textbf{System} &
      \makecell[r]{\textbf{Special}\\\textbf{HW/SW}} &
      \makecell[r]{\textbf{Patching}\\\textbf{Granularity}} &
      \makecell[r]{\textbf{Overhead}} &
      \makecell[r]{\textbf{Automotive}\\\textbf{Readiness}\\\textbf{(G1--G3)}} \\
      \midrule
      HERA       & Yes\tnote{1} & Instruction & Very low; static   & \texttimes \\
      RapidPatch & Yes\tnote{2} & Function    & Modest; dynamic  & \texttimes \\
      AutoPatch  & No           & Function    & Low; dynamic       & \texttimes \\
      Kintsugi   & Yes\tnote{3} & Instruction & Low; dynamic       & \texttimes \\
      \sname     & No           & Function    & Low; static        & $\checkmark$ \\
      \bottomrule
    \end{tabular}
    \begin{tablenotes}
      \item[1] \textit{FPB}: a hardware unit for debugging purposes.
      \item[2] \textit{eBPF}: a VM for running sandboxed code within the kernel.
      \item[3] \textit{Code-shadowing}: code is copied from flash to RAM for execution.
    \end{tablenotes}
  \end{threeparttable}
  \caption{High-level comparison of \sname with other hotpatching frameworks for real-time embedded systems.}
  \label{tab:hotpatch-compare}
\end{table}

\textbf{Hotpatching.} 
While hotpatching has been widely researched for general-purpose operating systems~\cite{windows_hotpatching, macos_hotpatching, arnold2009ksplice, su2024lpah, wang2023pet, xu2020automatic}, its application to real-time embedded systems is a more recent development. 
Several systems have been proposed for this domain. 
HERA~\cite{niesler2021hera} uses Flash Patch and Breakpoint (FPB) for instruction-level hotpatching. 
RapidPatch~\cite{he2022rapidpatch} utilizes Extended Berkeley Packet Filter (eBPF) to execute patch code in a sandboxed environment. 
AutoPatch~\cite{salehi2024autopatch} focuses on native patch execution and automatic patch generation from code commits. 
Kintsugi~\cite{kintsugi25} introduces a secure hotpatching solution for code-shadowing systems. 

Although these systems are important steps forward, they are fundamentally designed for general-purpose IoT devices and are not suited for the mission-critical automotive domain. 
As summarized in Table~\ref{tab:hotpatch-compare}, they all cannot achieve the three goals for automotive hotpatching. 
\myles{I significantly trimmed down the related hotpatching work. Do you think I need to bring back more details (e.g., providing some specific details about why they are not suitable for automotive systems)?}

\textbf{Conventional Update Mechanisms.} 
The most basic update method for embedded systems invovles erasing and reprogramming the firmware on flash memory~\cite{chlipala2004deluge}, which is not resilient to power loss and can easily "brick" a device. 
To mitigate this, some solutions employ a "minimal recovery kernel" that can run if an update fails~\cite{KACHMAN201991, software_updates_for_iot_ietf}. 
A more robust solution is A/B partitioning~\cite{android_ab_update, mcuboot_with_zephyr, freertos_ota_update_stm32u5}, where the system maintains two identical firmware partitions. 
Update is only applied to the inactive partition, which becomes active only after a successful verification and reboot. 
While A/B partitioning assures safe updates, it is costly, as it doubles the required flash storage, making it impractical for deeply embedded resource-constrained ECUs. 
Finally, there have been systems~\cite{bhatti2005mantis, vxworks_wind_river} that support dynamically loading code (i.e., loadable kernel modules), which works similarly to hotpatching, but it is complex and only provides limited coverage.

\textbf{Embedded File Systems.} 
Several embedded file systems~\cite{zephyr_vfs, FatFs, LittleFS, reliance-edge, tuxera-edgefs-cert} have been built to address the reliability issue of flash memory by employing copy-on-write (CoW) and/or transactional mechanisms. 
However, they are designed for general-purpose use and incur non-negligible performance and memory overheads. 
While a certified version like Tuxera EdgeFS CERT~\cite{tuxera-edgefs-cert} exists, it is proprietary, and does not come with the determinism and real-time safety required by us.  
\sname therefore realizes its own lightweight, "append-only" storage mechanism specifically designed for our use case.

%% file: contents/conclusion.tex
\section{Conclusion}

In this paper, we present \sname, the first hotpatching framework designed from the ground up for the stringent compliance, safety, and persistence requirements of automotive systems. 
It drastically reduces the MTTM for critical vulnerabilities and bugs. 
We achieve this by combining ASIL-D compliance readiness, bounded patching impact (both functional and timing-wise), and a corruption-resilient "append-only" flash storage mechanism. 
Our evaluation on automotive-grade hardware running both FreeRTOS and Zephyr OS confirms that \sname has a low overhead, deterministic behavior, and real-world applicability. 
The principles behind \sname are also applicable to other mission-critical domains, such as aerospace and medical devices, offering a path toward safer and more reliable software updates across the embedded systems landscape.

%% file: contents/appendices.tex
\cleardoublepage
\appendix
\section*{Ethical Considerations}
This paper is motivated by the need for a hotpatching solution for deeply embedded automotive systems, where a faster MTTM is highly desired. 
Given its characteristics, those systems are typically resource-constrained and safety-critical, and they are also highly regulated with standards such as ISO 26262. 
We present \sname, which is a framework designed with regulatory compliance in mind, and can provide safety-preserving and peristent hotpatching for those systems. 

\textbf{Beneficence.}
Our intent is for this work to yield better safety and security in automobiles, and we believe that \sname can help achieve this goal by providing a much faster MTTM. 
We also believe that \sname is generalizable to other mission-critical domains, which can help improve public safety. 
We do need car manufacturers and/or automotive ECUs vendors to support \sname in order to yield the expected benefits, which does need them to put in the development effort and cost, but we think \sname can help them reduce the development effort and cost for updates and provide better long-term support to their products; in addition, \sname can help them save cost on both flash memory (by not requiring A/B partitioning).
We also notice that if \sname is not implemented properly with respect to the required safety standard, it can yield unsafe behavior or introduce new vulnerabilities, which is why we have provided a detailed design guideline in this paper. 

\textbf{Respect for Persons.}
When designing \sname, we have always put compliance with the required safety standard as a top priority. 
When developing and evaluating \sname, we have conducted all the necessary tests and experiments on an automotive-grade hardware platform (but not in a real car). 
Using our prototype and evaluation of it, we have confirmed that \sname can yield the expected benefits. 
For the entire duration of this project, we did not collect or use any personal data, nor did we potentially endanger any person. 

\textbf{Justice.}
Even though we state that if \sname is not developed and integrated properly, it may yield unsafe behavior or introduce new vulnerabilities, we believe that the benefits upon a proper deployment of \sname are enormous, where it can significantly reduce the MTTM, which further means that automobile systems could be updated much faster (after a vulnerability or bug is found), thus improving public safety.
Therefore, this provides us the confidence that this paper is pro-justice, rather than anti-justice. 

\textbf{Respect for Law and Public Interest.}
As mentioned above (in Beneficence), we design \sname to be compliance-ready to the highest safety standard in automotive systems (i.e., ISO 26262 ASIL D). 
We also made sure to obey all relevant laws during the development and evaluation of \sname. 


\section*{Open Science}
\myles{Question for Sekar and Jorge: please check this section to make sure that it is also in compliance with Bosch Research's internal policy.}
We made sure that this paper is in full compliance with the Open Science policy. 
Our artifact of \sname, including its integration with FreeRTOS and Zephyr OS and evaluations with real-world CVEs, is made available at https://doi.org/10.5281/zenodo.18498465. 
As the hardware platform we use for evaluation contains some proprietary components from NXP, we cannot directly make them available; however, we notice that these proprietary components are available upon free license from NXP. 
We also intend to provide full documentation for \sname and its integration with FreeRTOS and Zephyr OS, as well as the evaluation of \sname with real-world CVEs, once the paper is published. 
\myles{Should I say this sentence or not? Technically we should make source available at the time of submission, but not its documentation; therefore, I'm thinking about make this statement here, though I'm a bit afraid that this might invite some criticism from the reviewers.}

\section{Placeholders For RTOS Task Functions}
\label{app:placeholder_4_task}

\begin{lstlisting}[language=C, style=VS2017, caption=C-style pseudo code of \sname placeholders for a RTOS task function., label={lst:placeholder_4_task}]
  extern uint32_t patch_readiness;
  extern uint32_t timing_buffers[32];

  void taskA(void* params)
  {
      // Variable definitions
      ...

      // Patchlings entry patch
      const uint32_t entry_timing_buffer = 15;
      uint8_t cur_task_pid = getPID();
      timing_buffers[cur_task_pid] = entry_timing_buffer;
      uint8_t call_status = PATCHLINGS_CALL_STATUS_NORMAL;
      if (patch_readiness >> current_task & 1)
          call_status = patchDispatcher(cur_task_pid, (void *)NULL);

      // Patchlings entry detour
      if (call_status == PATCHLINGS_CALL_STATUS_NORMAL)
      {
          // taskA entry code
          ...
      }

      // Patchlings entry timing delay
      patchlings_delay(cur_task_pid);
          
      // taskA main loop
      for (;;)
      {
          // Patchlings runtime patch
          const uint32_t runtime_timing_buffer = 10;
          timing_buffers[cur_task_pid] = runtime_timing_buffer;
          if (patch_readiness >> current_task & 1)
            call_status = patchDispatcher(cur_task_pid, (void *)NULL);

          // Patchlings runtime detour
          if (call_status == PATCHLINGS_CALL_STATUS_NORMAL) 
          {
              // taskA runtime code
              ...
          }
          else if (call_status == PATCHLINGS_CALL_STATUS_BREAK)
              break;

          // Patchlings runtime timing delay
          patchlings_delay(cur_task_pid);
      }

      // Patchlings exit patch
      const uint32_t exit_timing_buffer = 15;
      timing_buffers[cur_task_pid] = exit_timing_buffer;
      if (patch_readiness >> current_task & 1)
        call_status = patchDispatcher(cur_task_pid, (void *)NULL);

      // Patchlings exit detour
      if (call_status == PATCHLINGS_CALL_STATUS_NORMAL)
      {
          // taskA exit code
          ...
      }

      // Patchlings exit timing delay
      patchlings_delay(cur_task_pid);
  }
\end{lstlisting}

Here the Listing~\ref{lst:placeholder_4_task} demonstrate an example of how \sname placeholders are inserted into a RTOS task function (by default), along with the timing safety-preserving mechanism. 
We assume that there are at most 32 tasks in the system, and the timing buffer values are set to 15 $\mu$s and 10 $\mu$s for entry/exit and runtime, respectively.

\input{contents/security_considerations}

\begin{figure} 
    \centering
    \includegraphics[width=0.6\columnwidth]{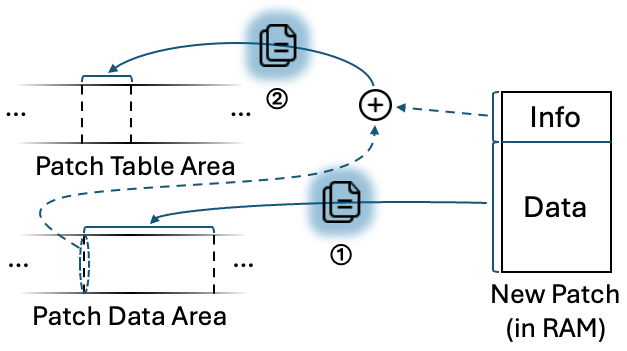}
    \caption{\em How \sname loads a patch to flash memory.}
    \label{fig:patch_loading}
\end{figure}

\section{Visualizing Patch Loading}
\label{app:visualizing_patch_loading}

We provide a visualization of how \sname loads a patch to flash memory in Figure~\ref{fig:patch_loading}. 

\begin{figure} 
    \centering
    \includegraphics[width=1\columnwidth]{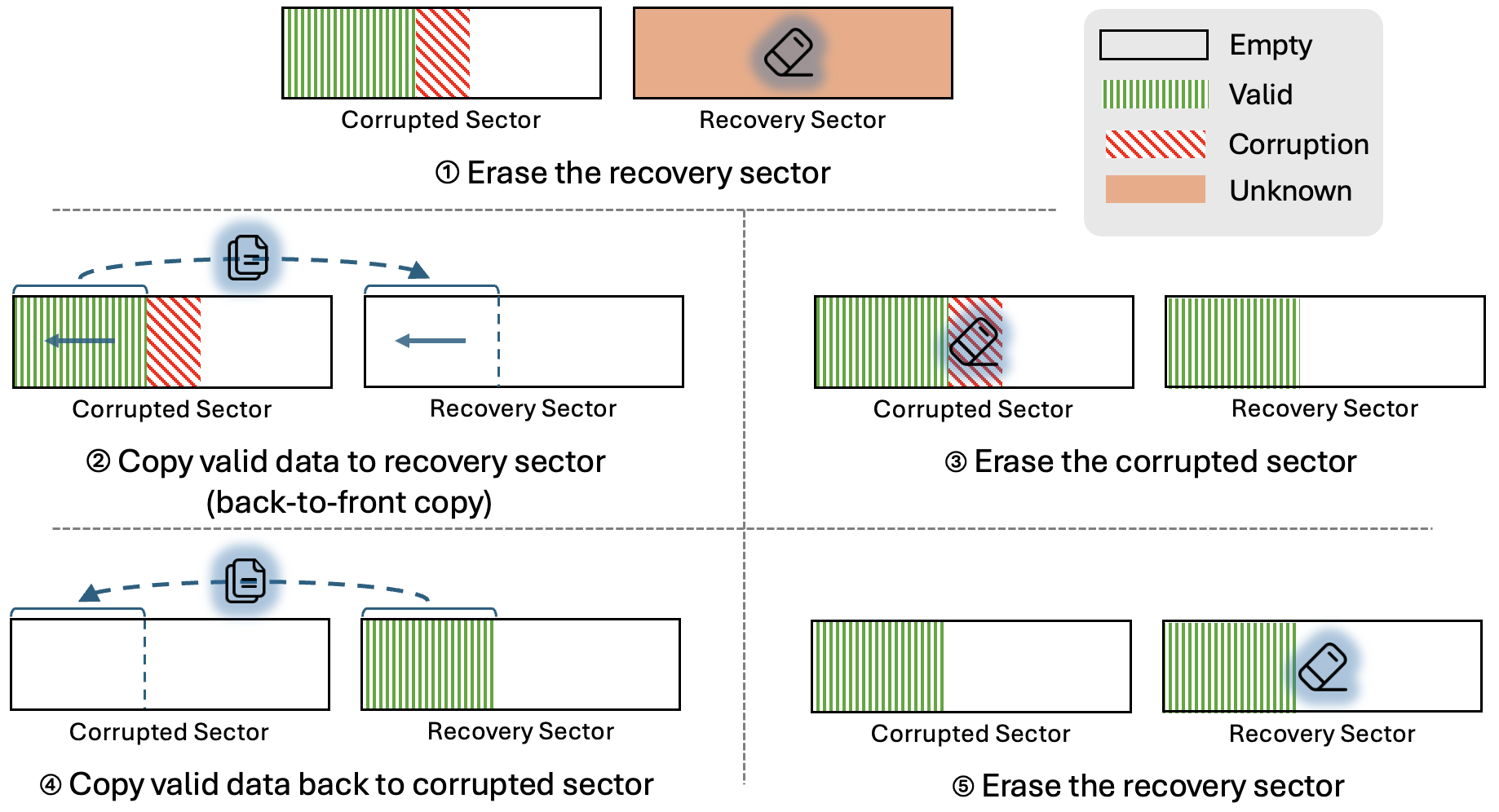}
    \caption{\em All steps of the recovery procedure in \sname.}
    \label{fig:recovery_procedure}
\end{figure}

\section{Technical Breakdown of Recovery Procedure}
\label{app:recovery_procedure}
Even with the "append-only" writing manner, flash corruption can still happen due to reasons such as power loss during flash writes. 
Thanks to the design of \sname's patch loading mechanism, the impact of flash corruption is limited to exactly one sector (i.e., the sector where the corruption happens). 

We define flash corruption into two types: dirty write and incomplete write. 
A dirty write happens when an active flash write operation is interrupted (e.g., power loss) before it is completed, causing the bits of the written area to be in an undefined state. 
An incomplete write happens when there is no active flash write operation, but the system is in the middle of writing a patch (as it involves multiple flash write operations to finish writing a patch) when a system reset happens, causing only part of the patch to be written to flash. 
In either case, the corrupted area becomes unusable. 
The difference is that in the case of a dirty write, the flash controller (with support of mechanisms such as ECC) usually gets involved to prevent any read to the corrupted area by issuing a bus fault, whereas in the case of an incomplete write, the corrupted area can still be read. 
We treat both cases equally as complete corruption of the affected storage area, and \sname employs a bus fault handler to catch any bus faults caused by dirty writes during flash reads, and uses the same recovery mechanism to handle both types of corruption.

To recover from a flash corruption, \sname first identify the sector where the corruption happens at boot time. 
It tries to read all \texttt{patch\_info} entries in the patch table one by one and check their validity.
If an invalid \texttt{patch\_info} is encountered (due to either flash corruption or missing data), \sname triggers the recovery procedure for the sector containing the invalid \texttt{patch\_info}. 
After the checks on the patch table are done, \sname proceeds to check if there is any non-empty data immediately after the last valid patch code, and if so, it also triggers the recovery procedure for the sector containing that data. 
Finally, if all checks pass, other \sname boot-time initialization steps are carried out. 

Once a corrupted sector is identified, \sname automatically triggers the recovery procedure. 
First, \sname erases the recovery sector and copy the still valid data from the corrupted sector to the recovery sector, in a back-to-front manner. 
This method of copying data ensures that even if a system reset happens during the recovery copy process, we can easily determine if the system is still in the recovery process during the next boot (to be discussed in the next paragraph). 
After the recovery copy is done, \sname then erases the corrupted sector and copy back all valid data from the recovery sector to the corrupted sector. 
Finally, the recovery sector is erased again to mark the end of the recovery process.
Figure~\ref{fig:recovery_procedure} demonstrates all the mentioned steps of the recovery procedure in \sname.

As mentioned above, the recovery process may be interrupted by a system reset, potentially causing data loss if not handled properly. 
To address this issue, at boot-time, \sname first checks if the beginning N bytes of the recovery sector contain any data, where N is the write unit size of the flash (i.e., the number of bytes that can be written to flash in a single write operation). 
Note that this is possible because of our back-to-front copy design mentioned above. 
If any data is found, it indicates that the system is in the middle of a recovery process, and \sname then resumes the recovery process. 
However, a new challenge arises here: how does \sname determine which sector is being recovered? 
There are essentially two cases here: a sector belonging to the patch table area is being recovered, or a sector belonging to the patch code area is being recovered.
In either case, the sector being recovered will be the immediate next sector after the last valid data in the respective area (i.e., patch table area or patch code area).
To determine which area is being recovered, \sname tries to read a patch table at the begining of the recovery sector (this is possible because \texttt{patch\_info} entries are aligned with flash sector boundaries). 
If a valid \texttt{patch\_info} is found, it indicates that a sector in the patch table area is being recovered; otherwise, it indicates that a sector in the patch code area is being recovered. 
In either case, there are two subcases: the corrupted sector still contains both valid and invalid data, or the corrupted sector is empty; the former subcase can be dealt with the same method we use to identify a corrupted sector, where the latter case is simply the immediate next sector after the last valid sector in the respective area (this is achievable because of our "append-only" writing manner). 
Finally, it continues the recovery process on the identified sector (starting from step 3 as illustrated in Figure~\ref{fig:recovery_procedure}) as described above until the recovery is completed. 

\section{Patching Assembly Code}
\label{app:patching_assembly_code}
By design, \sname only supports patching C/C++ code. 
On the other hand, assembly code exists in many embedded systems, especially in low-level system components such as bootloaders and hardware drivers. 
Patching assembly code is possible, and in fact, we implement preliminary support of patching assembly code and tested it for patching CVE-2020-10024 (a vulnerability related to performing Supervisor Call (SVC)) on Zephyr OS, where the vulnerable assembly code is patched successfully and the overall system overhead is similar (if not smaller) compared with patching C/C++ code.

However, assembly code lacks the safety mechanisms provided in high-level languages like C/C++, such as type safety and structured control flow, and the ASIL standards have strict verification requirements (especially for ASIL-D). 
To make matters worse, hotpatching assembly code can easily introduce hidden control and data flows that are hard to detect and verify, which can compromise the safety and reliability of the system.
Hence, we think that patching assembly code should be done with various formal analysis and verification techniques, which is out of the scope of this paper, and we leave proper support of it to future work.

\section{Global/Static Variables}
\label{app:global_static_variables}
\sname supports accessing and modifying global/static variables that already exist in the original program from a patch. 
However, if a patch wants to create its own global/static variables, there are currently two approaches, and both of them need to designate a reserved region in memory. 
One way is to compile all patches together and point all new global/static variables to that memory region, and another way is to directly make use of unused memory region with hard-coded addresses. 

%% file: contents/security_considerations.tex
\section{Security Considerations}
\label{app:security_considerations}
We focus heavily on compliance, safety, and persistence aspects of hotpatching in this paper, but it should be noted that security is also an important factor to consider.
Specifically, hotpatching, if not implemented carefully, can introduce new attack surfaces. 
Therefore, we propose several security considerations for designing and implementing a production ready version of \sname. 

\textbf{Integrity and Authenticity of Patches.}
Patches contain executable code, and as such, they should be protected against tampering. 
The integrity and authenticity of patches should be verified during the patch loading process by the \sname task using cryptographic certificates. 
Although we did not mention a specific design for this in this paper, we expect the patch certificate to be signed by either the ECU or car manufacturer, and the certificate can be stored as either part of the patch code or the patch's metadata (i.e., \texttt{patch\_info}), where the certificate authority used to verify the patch certificate should be pre-installed in the ECU by the car/ECU manufacturer.

\textbf{Control Flow Integrity.}
We also need to ensure that patches are executed in their desired contexts.
As the RTOS is considered as a trusted component, it is responsible for ensuring the control flow integrity of the ECU, even when an unprivileged task is compromised. 

\textbf{Isolation of The \sname Task.}
The \sname task is expected to be the sole task that is responsible for loading, verifying, and activating patches. 
This isolation is expected to be enforced by the RTOS, and the RTOS should prevent any other task from accessing the \sname task's memory space.
Also, to prevent existing patch data from being overwritten by other tasks, access control mechanisms should be in place to limit the access permission on region of address space of the flash memory controller (as the flash memory controller is the only component that is capable of writing or erasing data in the flash memory) to the \sname task, for example, by leveraging a Memory Protection Unit (MPU). 
In case of RAM-based hotpatching is enabled, the same isolation mechanism should be applied to the RAM-based patch storage region. 

\textbf{A Stronger Threat Model.}
In addition, if we assume a stronger threat model that physical attacks on the flash memory is possible, where the adversary may alter data stored in the flash memory, we expect two additional security considerations.
First, mechanisms such as secure boot should be used to protect both the original firmware and patch related data. 
Alternatively, if the secure boot mechanism cannot protect dynamically loaded patch related data, the ECU should conduct a complete authenticity and integrity verification of all patch related data before activating them. 
Second, additional runtime checks need to be conducted during the patch dispatching and execution process to prevent potential Time-Of-Check to Time-Of-Use (TOCTOU) attacks. 
However, we note that this check is a non-trivial workload, and even when a hardware accelerator is available, it may still add a moderate amount of delay to the patch dispatching and execution process.